\documentclass{aa}

\usepackage{graphicx}
\usepackage{placeins}
\usepackage{xcolor}

\usepackage{natbib,twoopt}
\usepackage[breaklinks=true, colorlinks=true,linkcolor=blue, citecolor=blue]{hyperref}
\bibpunct{(}{)}{;}{a}{}{,}             
\makeatletter
  \newcommandtwoopt{\citeads}[3][][]{\href{http://adsabs.harvard.edu/abs/#3}%
    {\def\hyper@linkstart##1##2{}%
     \let\hyper@linkend\@empty\citealp[#1][#2]{#3}}}
  \newcommandtwoopt{\citepads}[3][][]{\href{http://adsabs.harvard.edu/abs/#3}%
    {\def\hyper@linkstart##1##2{}%
     \let\hyper@linkend\@empty\citep[#1][#2]{#3}}}
  \newcommandtwoopt{\citetads}[3][][]{\href{http://adsabs.harvard.edu/abs/#3}%
    {\def\hyper@linkstart##1##2{}%
     \let\hyper@linkend\@empty\citet[#1][#2]{#3}}}
  \newcommandtwoopt{\citeyearads}[3][][]%
    {\href{http://adsabs.harvard.edu/abs/#3}
    {\def\hyper@linkstart##1##2{}%
     \let\hyper@linkend\@empty\citeyear[#1][#2]{#3}}}
\makeatother
\defcitealias{das2021}{Paper~I}
\usepackage{txfonts}
\begin{document}

   \title{Theoretical framework for BL~Her stars\\II. New period-luminosity relations in $Gaia$ passbands\thanks{Full Tables 2 and 4 are only available in electronic form at the CDS via anonymous ftp to \url{cdsarc.u-strasbg.fr (130.79.128.5)} or via \url{http://cdsweb.u-strasbg.fr/cgi-bin/qcat?J/A+A/}.}}


   \author{Susmita Das\inst{1,2},
           L\'aszl\'o Moln\'ar\inst{1,2,3},
            Shashi M. Kanbur\inst{4},
            Meridith Joyce\inst{1,2},
            Anupam Bhardwaj\inst{5},
          Harinder P. Singh\inst{6},
          Marcella Marconi\inst{5},
          Vincenzo Ripepi \inst{5}
            \and
            Radoslaw Smolec \inst{7}
          }

   \institute{Konkoly Observatory, HUN-REN Research Centre for Astronomy and Earth Sciences, Konkoly-Thege Mikl\'os \'ut 15-17, H-1121, Budapest, Hungary\\
              \email{susmita.das@csfk.org}
        \and
             CSFK, MTA Centre of Excellence, Budapest, Konkoly Thege Miklós út 15-17., H-1121, Hungary
        \and
            ELTE E\"otv\"os Lor\'and University, Institute of Physics and Astronomy, 1117, P\'azm\'any P\'eter s\'et\'any 1/A, Budapest, Hungary
        \and
            Department of Physics, State University of New York Oswego, Oswego, NY 13126, USA
         \and
             INAF-Osservatorio Astronomico di Capodimonte, Salita Moiariello 16, 80131, Naples, Italy
         \and
             Department of Physics \& Astrophysics, University of Delhi, Delhi 110007, India
         \and
            Nicolaus Copernicus Astronomical Center, Polish Academy of Sciences, Bartycka 18, PL-00-716 Warsaw, Poland
             }
\authorrunning{S. Das et al.}
   \date{Received 16 October 2023; accepted 21 January 2024}

 
  \abstract
     {In the era of the Hubble tension, it is crucial to obtain a precise calibration of the period-luminosity ($PL$) relations of classical pulsators. Type~II Cepheids (T2Cs; often exhibiting negligible or weak metallicity dependence on $PL$ relations) used in combination with RR~Lyraes and the tip of the red giant branch may prove useful as an alternative to classical Cepheids for the determination of extragalactic distances.}
   {We present new theoretical $PL$ and period-Wesenheit ($PW$) relations for a fine grid of convective BL~Her (the shortest period T2Cs) models computed using \textsc{mesa-rsp} in the $Gaia$ passbands and we compare our results with the empirical relations from $Gaia$ DR3. Our goal is to study the effect of metallicity and convection parameters on the theoretical $PL$ and $PW$ relations.}
   {We used the state-of-the-art 1D non-linear radial stellar pulsation tool \textsc{mesa-rsp} to compute models of BL~Her stars over a wide range of input parameters:\ metallicity ($-2.0\; \mathrm{dex} \leq \mathrm{[Fe/H]} \leq 0.0\; \mathrm{dex}$), stellar mass ($0.5M_{\odot}-0.8M_{\odot}$), stellar luminosity ($50L_{\odot}-300L_{\odot}$), and effective temperature (across the full extent of the instability strip; in steps of 50K). We used the Fourier decomposition technique to analyse the light curves obtained from \textsc{mesa-rsp} and $Gaia$ DR3 and then compared the theoretical and empirical $PL$ and $PW$ relations in the $Gaia$ passbands.}
   {The BL~Her stars in the All Sky region exhibit statistically different $PL$ slopes compared to the theoretical $PL$ slopes computed using the four sets of convection parameters. We find the empirical $PL$ and $PW$ slopes from BL~Her stars in the Magellanic Clouds to be statistically consistent with theoretical relations computed using the different convection parameter sets in the $Gaia$ passbands. There is a negligible effect coming from the metallicity on the $PL$ relations in the individual $Gaia$ passbands. However, there is a small but significant negative coefficient of metallicity in the $PWZ$ relations for the BL~Her models using the four sets of convection parameters. This could be attributed to the increased sensitivity of bolometric corrections to metallicities at wavelengths shorter than the $V$ band. Our BL~Her models also suggest a dependence of the mass-luminosity relation on metallicity. We found the observed Fourier parameter space to be covered well by our models. Higher mass models ($> 0.6\, M_{\odot}$) may be needed to reliably model the observed light curves of BL~Her stars in the All-Sky region. We also found the theoretical light curve structures (especially the Fourier amplitude parameters) to be affected by the choice of convection parameters.}
   {}


   \maketitle
%

\section{Introduction}

Type~II Cepheids (T2Cs) are low-mass (typically $< 1 M_{\odot}$) classical pulsators that are $1-4$ mag brighter than RR~Lyraes and $1.5-2$ mag fainter than classical Cepheids of similar periods. This makes them useful astrophysical objects, especially with respect to old, metal-poor stellar systems with scarce classical Cepheids and RR~Lyrae \citep[see reviews,][]{caputo1998, wallerstein2002, sandage2006}. T2Cs are known to pulsate in the fundamental mode \citep{bono1997a}. However, \citet{soszynski2019} recently discovered two first-overtone T2Cs in the Large Magellanic Cloud (LMC). Based on their pulsation periods, T2Cs are further classified as: the BL~Herculis stars (BL~Her: $1 \lesssim P \textrm{(days)}\lesssim 4$), the W~Virginis stars (W~Vir: $4 \lesssim P \textrm{(days)}\lesssim 20$), and the RV~Tauri stars (RV~Tau: $P \gtrsim 20$ days) \citep{soszynski2018}. We note that the period separation for the different sub-classes is not strict and may vary in different stellar environments. In addition, there is a fourth sub-class of brighter and bluer stars than W~Vir: the peculiar W~Vir (pW~Vir) stars \citep{soszynski2008}. Stars belonging to the different sub-classes exhibit different light curve morphologies and are expected to be at different evolutionary stages \citep[][]{gingold1985, wallerstein2002, bhardwaj2020a, bono2020}.

\citet{bono2020} noted that the evolutionary status of T2Cs is rather complex as compared to that of RR~Lyraes and classical Cepheids. The authors predicted BL~Her stars to be double-shell burning on the first crossing of the instability strip of the Hertzsprung-Russell diagram and W~Vir stars to be hydrogen-shell burning on the second crossing of the instability strip. This implies a common evolutionary channel for both the sub-types and the authors therefore suggest that BL~Her and W~Vir stars be considered a single group of classical pulsators. \citet{bono2020} also suggested a marginal dependence of the pulsation properties of T2Cs on metallicity.

 T2Cs obey well defined period-luminosity ($PL$) relations similar to RR~Lyraes (in the near-infrared bands) and classical Cepheids, which make them useful distance indicators \citep[see e.g.][]{ripepi2015, bhardwaj2017b, groenewegen2017b, braga2018, beaton2018b}. However, their more popular counterparts, namely, classical Cepheids, are more frequently studied and are known to exhibit strong metallicity effect on the $PL$ relations, especially at shorter wavelengths \citep[see, for example,][]{ripepi2021, ripepi2022, breuval2022, desomma2022, bhardwaj2023}. In contrast, T2Cs seem to exhibit weak or negligible metallicity effect on the $PL$ relations, as suggested by empirical \citep{matsunaga2006, matsunaga2009, matsunaga2011, groenewegen2017b,ngeow2022} and theoretical \citep{criscienzo2007, das2021} studies. T2Cs could therefore potentially exhibit a truly universal $PL$ relation throughout different stellar environments (and, thus, metallicities) and could serve as an alternative to classical Cepheids in the calibration of the extragalactic distance scale \citep{majaess2010,braga2020,das2021}. However, \citet{wielgorski2022} recently reported a significant metallicity effect on the near-infrared $PL$ relations of field T2Cs, although with the caveat that the sample size is rather small.

On the empirical front, stellar variability studies are undergoing a renaissance with ground-based wide-field time-domain surveys and space-based missions. In particular, the $Gaia$ mission \citep{prusti2016} promises to provide a percent-level calibration of the distance scale with parallaxes for a large sample of pulsating stars. With every subsequent data release \citep[DR1, DR2, EDR3, and DR3,][]{brown2016, brown2018, brown2021, vallenari2023}, the $Gaia$ mission provides multi-epoch observations and averaged astrophysical parameters for an increasing number of stars, with data for nearly 12 million variable sources in DR3 \citep[for a detailed summary of the properties of the different variable sources, see][]{eyer2023}. \citet{ripepi2023} presented the $Gaia$ DR3 catalogue of 15,006 Cepheids, including all subtypes, of which there are multi-band time-series photometry for 660 BL~Her stars. 

There are several linear and non-linear convective BL~Her models in the literature \citep{buchler1992, bono1995, bono1997a, bono1997b, marconi2007, smolec2012a, smolec2012b, smolec2014, smolec2016} but only a few that provide theoretical $PL$ and $PR$ relations for a large grid of BL~Her models \citep{criscienzo2007, marconi2007, das2021}. In a recent paper, \citet{das2021} (hereafter \citetalias{das2021}) computed a very fine grid of convective BL~Her models using the state-of-the-art 1D non-linear Radial Stellar Pulsation (\textsc{rsp}) tool within the \emph{Modules for Experiments in Stellar Astrophysics} \citep[\textsc{mesa},][]{paxton2011,paxton2013,paxton2015,paxton2018,paxton2019} software suite, encompassing a wide range of input parameters: metallicity ($-2.0\; \mathrm{dex} \leq \mathrm{[Fe/H]} \leq 0.0\; \mathrm{dex}$), stellar mass (0.5--0.8\,$M_{\odot}$), stellar luminosity (50--300\,$L_{\odot}$), and effective temperature (full extent of the instability strip; in steps of 50K). \textsc{mesa-rsp} is an open-source code that offers the possibility of generating light curves of classical pulsators at multiple wavelengths, while also varying the convective parameters. Although the longer-period T2Cs pose challenges for the existing pulsation codes due to their highly non-adiabatic nature, BL~Her stars may be modelled reliably using \textsc{mesa-rsp}. The higher mass models were inspired by the grid of $0.6\,M_{\odot}$ and $0.8\,M_{\odot}$ non-linear convective T2C models in \citet{smolec2016}. Furthermore, \citetalias{das2021} obtained theoretical $PL$ and $PR$ relations for BL~Her models at the Johnson-Cousins-Glass bands $UBVRIJHKLL'M$ and tested the effects of metallicity and convection parameters on these relations.

A comparison of the theoretical mean light relations and light curves of classical pulsators with those from observations not only provides stringent constraints for the stellar pulsation models, but also serves as probes for understanding the theory of stellar evolution and pulsation \citep{simon1985, marconi2013a, marconi2017, bhardwaj2017a, das2018, das2020}. Of noteworthy importance is the case of period doubling in BL~Her stars, which was predicted by \citet{buchler1992} and was later observed (and consistently modelled) in a 2.4-d BL~Her type variable in the Galactic bulge \citep{soszynski2011, smolec2012a}. \citetalias{das2021} compared theoretical $PL$ relations for BL~Her models computed using \textsc{mesa-rsp} with empirical $PL$ relations in the LMC \citep{matsunaga2009, bhardwaj2017b, groenewegen2017b}, SMC \citep{matsunaga2011, groenewegen2017b}, Galactic bulge \citep{bhardwaj2017c}, and Galactic globular clusters \citep{matsunaga2006}, along with prior theoretical $PL$ relations derived by \citet{criscienzo2007}. More recently, \citet{jurkovic2023} presented results from T2Cs in the $Kepler$ K2 mission and corresponding BaSTI (Bag of Stellar Tracks and Isochrones) evolutionary models \citep{hidalgo2018} with metallicities, $\mathrm{[Fe/H]} = -2.50, -1.20$ and $0.06$, and masses from 0.45\,$M_{\odot}$ to 0.8\,$M_{\odot}$. 

As an extension to \citetalias{das2021}, the aim of this project is to (i) compute $Gaia$ passband light curves for a very fine grid of BL~Her models encompassing a wide range of metallicities, stellar mass, stellar luminosity and effective temperature using four different sets of convection parameters with \textsc{mesa-rsp}, (ii) obtain new theoretical $PL$ and $PW$ relations in the $Gaia$ passbands, (iii) study the effect of metallicity and convection parameters on these relations, and (iv) identify whether there exists any particular convection set that is preferred by the empirical results. The structure of the paper is as follows: the theoretical and empirical data used in this analysis are described in Section~\ref{sec:data}. We present new theoretical $PL$ and $PW$ relations of the BL~Her models in the $Gaia$ passbands in Section~\ref{sec:PL} and investigate the effect of metallicity and convection parameters. We probe the dependence of the mass-luminosity relation of the BL~Her models on the chemical composition in Section~\ref{sec:ml}. We compare the theoretical and observed Fourier parameters of the BL~Her light curves in Section~\ref{sec:FP}.  Finally, we summarize the results of our study in Section~\ref{sec:results}.

\section{Data and methodology}
\label{sec:data}
\subsection{Stellar pulsation models}

A detailed description of the BL~Her models computed using \textsc{mesa-rsp} and used for this analysis is presented in \citetalias{das2021}. In brief, we computed a fine grid of BL~Her models using four different sets of convection parameters \citep[see, Table~4 of][]{paxton2019}, each with increasing complexity: set~A (the simplest convection model), set~B (radiative cooling added), set~C (turbulent pressure and turbulent flux added), and set~D (all of these effects added simultaneously). The grid of stellar pulsation models computed using \textsc{mesa-rsp} (i) uses the \citet{kuhfuss1986} theory of turbulent convection (ii) follows the treatment of stellar pulsation as outlined in \citet{smolec2008}, and (iii) takes into account the structure of the non-rotating and chemically homogeneous stellar envelope only, without considering the detailed core structure. The BL~Her models were computed using \textsc{mesa} version r$11701$\footnote{The \textsc{mesa} inlists used for computing the BL~Her models in the $Gaia$ passbands may be found in \url{https://doi.org/10.5281/zenodo.10404005}}.

\subsection{Linear and non-linear computations}

\begin{table}
\caption{Chemical compositions of the adopted pulsation models.}
\centering
\begin{tabular}{c c c}
\hline
[Fe/H] & $Z$ & $X$\\
\hline
-2.00 & 0.00014 & 0.75115\\
-1.50 & 0.00043 & 0.75041\\
-1.35 &0.00061& 0.74996\\
-1.00 &0.00135& 0.74806\\
-0.50&0.00424&0.74073\\
-0.20 &0.00834& 0.73032\\
0.00 &0.01300& 0.71847\\
\hline
\end{tabular}
\tablefoot{\small
	 The $Z$ and $X$ values are estimated from the $\mathrm{[Fe/H]}$ values by assuming the primordial helium value of 0.2485 from the WMAP CMB observations \citep{hinshaw2013} and the helium enrichment parameter value of 1.54 \citep{asplund2009}. The solar mixture is adopted from \citet{asplund2009}.}
\label{tab:composition}
\end{table}

We begin with the linear computations for a fine grid of BL~Her models with the following input parameters for each of the four convection parameter sets:

\begin{enumerate}
\item Metallicity (see Table~\ref{tab:composition} for corresponding $ZX$ values)

$\mathrm{[Fe/H]} = -2, -1.5, -1.35, -1, -0.5, -0.2, 0$

\item Stellar mass ($M$)
\begin{enumerate}
\item Low-mass range = $0.5\,M_{\odot}$, $0.55\,M_{\odot}$, $0.6\,M_{\odot}$
\item High-mass range = $0.65\,M_{\odot}$, $0.7\,M_{\odot}$, $0.75\,M_{\odot}$, $0.8\,M_{\odot}$
\end{enumerate}

\item Stellar luminosity ($L$)
\begin{enumerate}
\item For low-mass range = $50-200\,L_{\odot}$, in steps of 50\,L$_{\odot}$
\item For high-mass range = $50-300\,L_{\odot}$, in steps of 50\,L$_{\odot}$
\end{enumerate}

\item Effective temperature ($T_{\rm{eff}}$) =  $4000-8000$\,K, in steps of 50\,K.
\end{enumerate}

Each $ZXMLT_\textrm{eff}$ combination serves as the input stellar parameters for the linear computations; equilibrium static models are constructed, and their linear stability analysis conducted, resulting in linear periods of the radial pulsation modes and their respective growth rates \citep{paxton2019}. Positive growth rates in the fundamental mode help in estimating the boundaries of the IS (see Fig.~1 of \citetalias{das2021}). The equilibrium static models from the linear computations serve as input for the non-linear model integration. The non-linear model integrations are carried for 4000 pulsation cycles each and only models with non-linear periods between $1-4$ days are considered as BL~Her models, following the conventional classification for BL~Her stars \citep{soszynski2018}. Finally, we check for single periodicity and full-amplitude stable pulsation of the models using the condition that the fractional growth of the kinetic energy per pulsation period $\Gamma$ does not vary by more than 0.01, the amplitude of radius variation $\Delta R$ by more than 0.01~$R_{\odot}$, and the pulsation period $P$ by more than 0.01~d over the last 100 cycles of the completed integration. The total number of BL~Her models in each convection parameter set finally accepted for this analysis is summarised in Table~\ref{tab:number}.

BL~Her stars are traditionally considered to be low mass ($M\,\leq\, 0.6\,M_{\odot}$) stars, which evolve towards the Asymptotic Giant Branch (AGB) from the blue edge of the Zero Age Horizontal Branch (ZAHB), crossing the instability strip at luminosities higher than RR~Lyrae stars \citep{gingold1985, bono1997a, caputo1998}. However, a photometric study of the prototype, BL~Herculis itself, by \citet{alexander1987} estimated its stellar mass to be $M\,\approx\, 0.75\,M_{\odot}$. A few theoretical studies for BL~Her stars also include masses higher than $0.6\,M_{\odot}$ \citep{marconi2007, smolec2012a}. Furthermore, a wider mass range can accommodate the differences between evolutionary codes, as these have uncertainties of their own depending on the assumptions they include, e.g., on convection efficiency. In light of this, we use the stellar masses of BL~Her models from evolutionary tracks not as stringent constraints but as useful hints towards our choice of input parameters in the pulsation code and explore the possibility of higher mass models. However, for a reliable comparison with observations, we carry out the theoretical analysis twice: first, with the entire set of BL~Her models (low mass+high mass; $0.5-0.8\,M_{\odot}$) and next, with the subset of low mass BL~Her models ($0.5-0.6\,M_{\odot}$) only. We note here that the higher stellar mass ($M\,> 0.6\,M_{\odot}$) and lower metallicity ($Z=0.00014$) models may have an overlap with evolved RR~Lyraes. A detailed theoretical investigation is crucial to shed light on the longstanding RR~Lyrae--T2C separation problem based on their evolutionary status \citep[see also,][]{braga2020}.

\subsection{Processing the model data}

\begin{table*}
\caption{Light curve parameters of the BL~Her models used in this analysis computed using \textsc{mesa-rsp}. The columns provide the chemical composition ($ZX$), passband ($\lambda$), stellar mass ($\frac{M}{M_{\odot}}$), mode of pulsation, effective temperature ($T_{\rm eff}$), stellar luminosity ($\frac{L}{L_{\odot}}$), logarithmic pulsation period ($\log(P)$), amplitude ($A$), mean magnitude ($m_0$), Fourier amplitude ($R_{21}$,$R_{31}$) and phase ($\phi_{21}$,$\phi_{31}$) parameters, mean radius ($\log\frac{R}{R_{\odot}}$), and the convection parameter set used.}
\centering
\scalebox{0.85}{
\begin{tabular}{c c c c c c c c c c c c c c c c}
\hline\hline
$Z$ & $X$ & $\lambda$ & $\frac{M}{M_{\odot}}$ & Mode & $T_{\rm eff}$ & $\frac{L}{L_{\odot}}$ & $\log(P)$ & $A$ & $m_0$ & $R_{21}$ & $R_{31}$ & $\phi_{21}$ & $\phi_{31}$ & $\log\frac{R}{R_{\odot}}$ & Convection set\\
[0.5ex]
\hline \hline
0.01300&        0.71847&        $G$&    0.50&   FU&     5900&   50&     0.036&  0.068&  0.377&  0.288&  0.046&  4.672&  1.048&  0.834&  A\\
0.01300&        0.71847&        $G$&    0.50&   FU&     5950&   50&     0.022&  0.104&  0.375&  0.494&  0.103&  4.565&  1.029&  0.827&  A\\
...&    ...&    ...&    ...&    ...&    ...&    ...&    ...&    ...&    ...&    ...&    ...& ...&    ...&    ...&    ...\\
\\
0.01300&        0.71847&        $G$&    0.50&   FU&     5900&   50&     0.038&  0.110&  0.379&  0.390&  0.103&  4.610&  1.028&  0.836&  B\\
0.01300&        0.71847&        $G$&    0.50&   FU&     5950&   50&     0.024&  0.156&  0.376&  0.592&  0.178&  4.634&  1.110&  0.829&  B\\
...&    ...&    ...&    ...&    ...&    ...&    ...&    ...&    ...&    ...&    ...&    ...& ...&    ...&    ...&    ...\\
\\
0.01300&        0.71847&        $G$&    0.50&   FU&     5650&   50&     0.124&  0.009&  0.406&  0.020&  0.004&  5.247&  1.976&  0.888&  C\\
0.01300&        0.71847&        $G$&    0.50&   FU&     5700&   50&     0.106&  0.160&  0.402&  0.367&  0.147&  5.179&  2.032&  0.879&  C\\
...&    ...&    ...&    ...&    ...&    ...&    ...&    ...&    ...&    ...&    ...&    ...& ...&    ...&    ...&    ...\\
\\
0.01300&        0.71847&        $G$&    0.50&   FU&     5700&   50&     0.107&  0.139&  0.402&  0.431&  0.185&  5.278&  2.025&  0.880&  D\\
0.01300&        0.71847&        $G$&    0.50&   FU&     5750&   50&     0.090&  0.170&  0.397&  0.650&  0.305&  5.434&  2.201&  0.871&  D\\
...&    ...&    ...&    ...&    ...&    ...&    ...&    ...&    ...&    ...&    ...&    ...& ...&    ...&    ...&    ...\\
\\
\hline
\end{tabular}}
\tablefoot{\small 
        This table is available entirely in electronic form at the CDS.}
\label{tab:allmodels}
\end{table*}

\begin{figure}
\centering
\includegraphics[scale = 0.95]{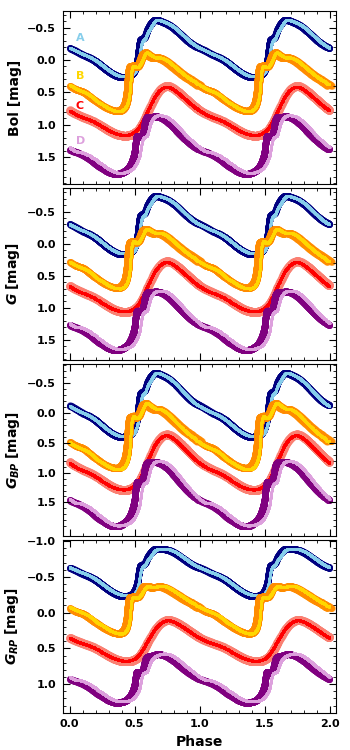}
\caption{Light curves in the bolometric, $G$, $G_{BP}$, and $G_{RP}$ bands (shown from top to bottom) for a BL~Her model with input parameters, $Z=0.00043$, $X=0.75041$, $M=0.6\,M_{\odot}$, $L=100\,L_{\odot}$, and $T=6700$K computed using four sets of convection parameters (sets A, B, C, and D depicted in different colours). Light curves are vertically offset by 0.5 mag successively for visualisation purposes. The light curves in the foreground are computed using \textsc{mesa}~r11701 and used in this analysis while the corresponding light curves depicted in the background are computed using the latest version, \textsc{mesa}~r23.05.1.}
\label{fig:LC}
\end{figure}

One of the outputs of the \textsc{mesa-rsp} non-linear model computations is bolometric light curves, which can thereby be transformed into light curves in the required passbands, as described in \citet{paxton2018} and \citetalias{das2021}. The transformation of the absolute bolometric magnitudes into the absolute magnitudes in the Johnson-Cousins-Glass bands $UBVRIJHKLL'M$ in \citetalias{das2021} was done using pre-computed bolometric correction tables from \citet{lejeune1998} included in \textsc{mesa}. However, pre-computed bolometric correction tables to transform bolometric light curves into $Gaia$ passbands are not yet included in \textsc{mesa}. We therefore opted to include user-provided bolometric correction tables defined as a function of the stellar photosphere in terms of $T_{\rm eff}$\,(K), $\log (g ({\rm cm \; s^{-2}}))$, and metallicity [M/H]. The bolometric correction tables used to transform the theoretical bolometric light curves into the $Gaia$ passbands ($G$, $G_{BP}$, and $G_{RP}$) in this study are obtained from the MIST\footnote{\url{https://waps.cfa.harvard.edu/MIST/index.html}} (\textsc{mesa} Isochrones \& Stellar Tracks) packaged model grids; the bolometric correction tables were computed using 1D atmosphere models \citep[based on ATLAS12/SYNTHE;][]{kurucz1970, kurucz1993}. An example of the light curves in the bolometric and the Gaia passbands ($G$, $G_{BP}$, and $G_{RP}$) for a BL~Her model computed using the four sets of convection parameters is presented in Fig.~\ref{fig:LC}. 

As mentioned earlier, the BL~Her models used in this work were computed using \textsc{mesa}~r11701; however, the latest version of the software is now \textsc{mesa}~r23.05.1. To test the compatibility of our models with the latest \textsc{mesa} version, we computed a particular BL~Her model with the input parameters, $Z=0.00043$, $X=0.75041$, $M=0.6\,M_{\odot}$, $L=100\,L_{\odot}$, and $T=6700$K using \textsc{mesa}~r23.05.1. The comparison of the light curves of the BL~Her model in the $Gaia$ passbands and over the four sets of convection parameters computed using both \textsc{mesa} versions is displayed in Fig.~\ref{fig:LC}. We find the outputs from both \textsc{mesa} versions to be quite similar; this is because while there have been changes incorporated in the \textsc{mesa} software itself, resulting in the release of different versions\footnote{For a detailed list of the changes in the different versions of \textsc{mesa}, the interested reader is referred to \url{https://docs.mesastar.org/en/release-r23.05.1/changelog.html}}, no significant changes have been made to the \textsc{mesa-rsp} module of the software. We note that while \textsc{mesa} is a 1D stellar evolution code, we have made use of the \textsc{mesa-rsp} code for this work, which is responsible for the computations of non-linear radial stellar pulsations.

As in \citetalias{das2021}, the theoretical light curves in the $Gaia$ passbands are fitted with the Fourier sine series \citep[see, for example,][]{deb2009, das2018} of the form:
\begin{equation}
m(x) = m_0 + \sum_{k=1}^{N}A_k \sin(2 \pi kx+\phi_k),
\label{eq:fourier}
\end{equation}

\noindent where $x$ is the pulsation phase, $m_0$ is the mean magnitude, and $N$ is the order of the fit ($N = 20$). In this work, in addition to deriving new theoretical $PL$ and $PW$ relations using $m_0$, as was done in \citetalias{das2021}, we also study the light curve structures using the Fourier decomposition technique. The Fourier amplitude and phase coefficients ($A_k$ and $\phi_k$) have been re-defined as Fourier parameters in terms of amplitude ratios and phase differences as follows:
\begin{equation}
\begin{aligned}
R_{k1} &= \frac{A_k}{A_1}, \\
\phi_{k1} &= \phi_k - k\phi_1,
\label{eq:params}
\end{aligned}
\end{equation}
\noindent where $k > 1$ and $\rm 0 \leq \phi_{k1} \leq 2\pi$. The errors in the Fourier parameters are thereby calculated using error propagation in the Fourier coefficients \citep[see][]{deb2010}. Table~\ref{tab:allmodels} summarises the input stellar parameters of the BL~Her models used in this analysis, along with the computed light curve parameters in the different $Gaia$ passbands obtained from the Fourier fitting.

We also defined the peak-to-peak amplitude $A$ of the light curve as the difference in the minimum and the maximum of the light variations:
\begin{equation}
    A_{\lambda} = (M_{\lambda})_{\rm min} - (M_{\lambda})_{\rm max},
\end{equation}
where $(M_{\lambda})_{\rm min}$ and $(M_{\lambda})_{\rm max}$ are the minimum and maximum magnitudes obtained from the Fourier fits in the $\lambda$ band, respectively. 

\subsection{Observations from $Gaia$ DR3}

The third data release of the European Space Agency's (ESA) $Gaia$ mission \citep[$Gaia$ DR3;][]{prusti2016, vallenari2023} is based on data collected over a period of 34 months and provides an unprecedented wealth of epoch photometry in the $G$, $G_{BP}$, and $G_{RP}$ passbands, among other astrophysical parameters. One important difference of $Gaia$ DR3 from $Gaia$ DR2 is the existence of five sub-regions in $Gaia$ DR3: the Large and Small Magellanic Clouds (LMC and SMC), the Andromeda (M31) and Triangulum (M33) galaxies, and all of the remaining sky excluding the four sub-regions (referred to as All Sky), as opposed to three sub-regions in $Gaia$ DR2. For more details on how these sub-regions are defined, the interested reader is referred to Table~1 of \citet{ripepi2023}.

The $Gaia$ DR3 catalogue contains a sample of 15,006 Cepheids of all types across the five sub-regions. However, the completeness of the sample varies significantly with the region of the sky and also with the Cepheid type \citep{ripepi2023}. In particular, owing to their faint magnitudes, no BL~Her star has been detected in the distant M31 and M33 galaxies. This study therefore includes observed BL~Her stars in the LMC, SMC and All Sky sub-regions of $Gaia$ DR3 only. 
At first glance, the $Gaia$ DR3 catalogue contains 545 BL~Her stars. However, care should be taken to take objects incorrectly classified by the Specific Objects Study (SOS) Cep\&RRL pipeline in $Gaia$ DR3 into account. A straightforward cross-match between the $Gaia$ DR3 catalogue and Table~6 of \citet{ripepi2023} adds 221 more BL~Her stars while removing 106 misclassified BL~Her stars, resulting in a total of 660 BL~Her stars in the LMC, SMC and All Sky regions from $Gaia$ DR3\footnote{The observed data from $Gaia$ DR3 used in this work are available at \url{https://gea.esac.esa.int/archive/}}. Table~\ref{tab:Observed_Data} presents the number of BL~Her stars with photometric data in the three $Gaia$ passbands. 

In an effort to make use of the best-quality observed light curves available for BL~Her stars from the the $Gaia$ DR3 catalogue, we restricted our sample to include only those stars with more than 20 epoch observations for further analysis. As with the theoretical BL~Her light curves, the observed light curves of the BL~Her stars in the $Gaia$ passbands from $Gaia$ DR3 are also fitted with the Fourier sine series of the mathematical form as given in Eq.~\ref{eq:fourier}. We first fit the observed light curves with an initial order of fit $N = 4$ and remove the points lying $2\sigma$ away from the fit. We then use the Baart's condition \citep{baart1982} to obtain the optimum order of fit by varying $N$ from 4 to 8, as described below \citep[see, for more details,][]{petersen1986}.

\begin{table}
\caption{Number of observed BL~Her stars with epoch photometry data in the $Gaia$ passbands from the $Gaia$ DR3 catalogue in the All Sky, LMC, and SMC sub-regions. The last column presents the number of stars with observations having more than 20 epochs and finally used in this analysis.}
\centering
\begin{tabular}{c c c c }
\hline\hline
Sub-region & N$_\textrm{BL~Her}$ & Band & N$_\textrm{BL~Her}$\\
 &  & & (Finally used)\\
\hline \hline
& & $G$ & 430\\
All Sky & 579 & $G_{BP}$ & 412\\
& & $G_{RP}$ & 417\\
\hline
& & $G$ &65\\
LMC & 65 & $G_{BP}$ & 58\\
& & $G_{RP}$ &  58\\
\hline
& & $G$ & 16\\
SMC & 16 & $G_{BP}$ & 14\\
& & $G_{RP}$ & 14\\
\hline
\end{tabular}
\label{tab:Observed_Data}
\end{table}

\begin{table*}
\caption{Light curve parameters of observed BL~Her stars. The columns provide the sub-region, passband ($\lambda$), $Gaia$ Source ID, logarithmic period, order of Fourier fit, amplitude ($A$), mean magnitude ($m_0$), and the Fourier amplitude ($R_{21}$,$R_{31}$) and phase ($\phi_{21}$,$\phi_{31}$) parameters.}
\centering
\scalebox{0.95}{
\begin{tabular}{c c c c c c c c c c c}
\hline\hline
Sub-region & $\lambda$ & $Gaia$ Source ID & $\log(P)$ & Order & A  & $m_0$ & $R_{21}$ & $R_{31}$ & $\phi_{21}$ & $\phi_{31}$\\
& & & & & & $\sigma_{m_0}$ & $\sigma_{R_{21}}$ & $\sigma_{R_{31}}$ & $\sigma_{\phi_{21}}$ & $\sigma_{\phi_{31}}$\\
[0.5ex]
\hline \hline
All Sky&        $G$&    4116810545125634432&    0.535&  7&      0.677&  15.234&         0.262&          0.100&          3.379&  0.173   \\
& & & & & & 0.002& 0.006 & 0.006& 0.032&        0.073\\
All Sky&        $G$&    4116883452108512128&    0.302&  4&      0.523&  16.635& 0.214&          0.097&          3.904&  5.814   \\
& & & & & & 0.009& 0.050& 0.050& 0.300& 0.622\\
...&    ...&    ...&    ...&    ...&    ...&    ...&    ...&    ...&    ...& ...\\
LMC&    $G$&    4658053557649384320&    0.265&  4&      0.454&  18.378& 0.032&          0.190&          2.863   &       4.790   \\
& & & & & & 0.004& 0.025&0.027&0.977&0.169\\
LMC&    $G$&    4658088604613882752&    0.345&  4&      0.377&  17.878&         0.405&  0.111&  2.532   &5.236  \\
& & & & & & 0.003&0.033 &0.020  &0.064  &0.215\\
...&    ...&    ...&    ...&    ...&    ...&    ...&    ...&    ...&    ...& ...\\
SMC&    $G$&    4685850311156320640&    0.173&  6&      0.270&  18.657& 0.136   & 0.147   & 3.131 & 5.690 \\
& & & & & & 0.006       & 0.057 & 0.073 & 0.498 & 0.377\\
SMC&    $G$&    4685994518965321088&    0.274&  4&      0.791&  18.189& 0.373&0.198     &2.608  &5.488  \\
& & & & & & 0.011       & 0.011 & 0.011 & 0.163 & 0.267\\ 
...&    ...&    ...&    ...&    ...&    ...&    ...&    ...&    ...&    ...& ...\\
\hline
\end{tabular}}
\tablefoot{\small 
        This table is available entirely in electronic form at the CDS.}
\label{tab:Observed}
\end{table*}

Care should be taken to use the optimum number of terms in the Fourier decomposition of the observed light curves. This is because an $N$ too small results in systematic deviations from the best estimate while an $N$ too large will also fit the noise. The optimum order of fit is the one for which the residuals of the fit consist of noise only, without any trends from the residuals. To determine this, we used the unit-lag auto-correlation of the sequence of residuals \citep{baart1982, petersen1986} defined as:

\begin{equation}
    \rho:= \frac{\sum_{j=1}^{n}(v_j-\bar{v})(v_{j+1}-\bar{v})}{\sum_{j=1}^{n}(v_j-\bar{v})^2},
   \label{eq:rho} 
\end{equation}

where $v_j$ is the $j$th residual, $\bar{v}$ denotes the average of the residuals and $j$ = 1, ...n represents the number of data points in the observed light curve. The residual of the fitted light curve $v$ is obtained as follows:

\begin{equation}
    v = m(x) - [m_0 + \sum_{k=1}^{N}A_k \sin(2 \pi kx+\phi_k)].
\end{equation}

For the calculation of $\rho$, we chose the ordering of $v_j$ not from the original time-sequence but from the order of increasing phase values. Residuals with a definite trend will result in $\rho \approx 1$. Uncorrelated residuals will exhibit smaller values of $\rho$, with the idealized case of residuals with equal magnitudes and alternating signs resulting in $\rho \approx -1$. According to \citet{baart1982}, a value of $\rho \geq [n-1]^{-1/2}$ ($n$ is the number of observations) indicates  the presence of a trend, while $\rho \leq 2[n-1]^{-1/2}$ suggests a lack of any trend in the residuals. The auto-correlation cut-off tolerance is therefore defined as:

\begin{equation}
    \rho_c:= \rho(cut) = [2(n-1)]^{-1/2}.
\end{equation}

The optimum order of fit $N$ is thereby obtained as follows: a series of Fourier decompositions is calculated using Eq.~\ref{eq:fourier} with increasing $N$ from 4 to 8. For every decomposition, $\rho=\rho(N)$ was calculated using Eq.~\ref{eq:rho}. The value of $N$ for which $|\rho-\rho_c|$ is minimised is the one for which residuals consist of noise only and is chosen as the optimum order of fit.

Finally, we fit the observed light curves with outliers $\geq 2\sigma$ removed using the optimum order of fit. The Fourier parameters and associated errors are thereby calculated using Eq.~\ref{eq:params}. The number of observed BL~Her stars in the respective $Gaia$ passbands that were ultimately used in this analysis is summarised in the last column of Table~\ref{tab:Observed_Data} and the light curve parameters of the observed BL~Her stars are listed in Table~\ref{tab:Observed}.

\section{Period-luminosity relations}
\label{sec:PL}

The mean magnitudes obtained from the Fourier-fitted observed and theoretical light curves of the BL~Her stars are used to obtain period-luminosity ($PL$) relations in the $Gaia$ passbands ($G$, $G_{BP}$, $G_{RP}$) of the mathematical form:
\begin{equation}
\begin{aligned}
M_\lambda =& a\log(P)+b \quad \textrm{(for \textsc{mesa-rsp} BL~Her models)}; \\
m_\lambda =& a\log(P)+b \quad \textrm{(for $Gaia$ DR3 BL~Her stars)},
\end{aligned}
\end{equation}
where $M_\lambda$ and $m_\lambda$ refer to the absolute and the apparent magnitudes in a given band, $\lambda$, respectively. We note here that the apparent magnitudes of the observed BL~Her stars from $Gaia$ DR3 have not been corrected for extinction in the present study. This follows from \citet{ripepi2019}, where the authors noted that extinction only affects the $PL$ zero points, while keeping the $PL$ slopes unaffected in LMC and SMC, given their small average foreground reddening values of the order of $E(B-V) \approx$  0.08 and 0.04 mag, respectively.

However, to account for uncertainties related to $PL$ relations obtained using apparent magnitudes without extinction correction, we also use Wesenheit indices \citep{madore1982}, which act as pseudo-magnitudes and thereby derive theoretical and empirical period-Wesenheit ($PW$) relations. We use the new Wesenheit magnitudes as defined by \citet{ripepi2019} for $Gaia$ DR3:
\begin{equation}
 W(G, G_{BP} - G_{RP})=G - 1.90(G_{BP}-G_{RP}), 
 \label{eq:wesenheit}
\end{equation}
where $G$, $G_{BP}$, and $G_{RP}$ are the apparent and the absolute magnitudes in the respective $Gaia$ passbands for the empirical and theoretical $PW$ relations, respectively, of the mathematical form:
\begin{equation}
W = a\log(P)+b.
\end{equation}

\begin{figure*}
\centering
\includegraphics[scale = 0.95]{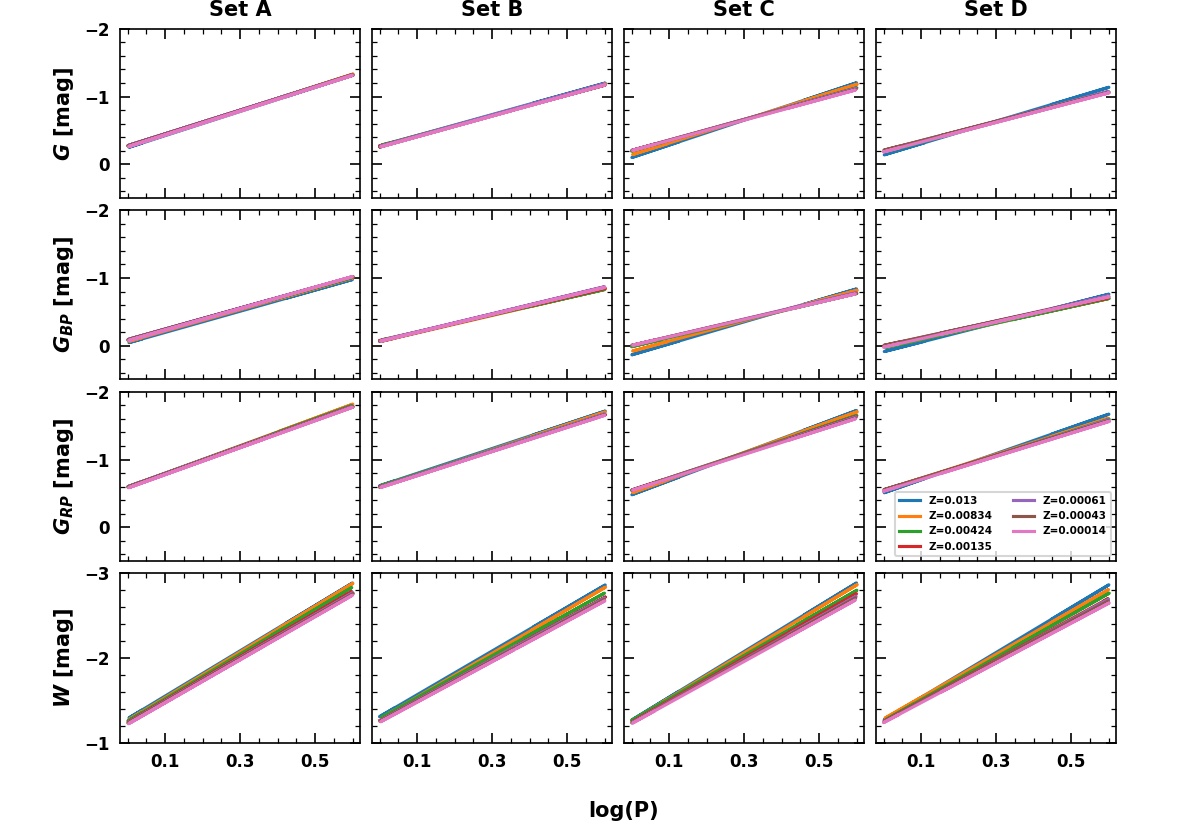}
\caption{$PL$ and $PW$ relations of the BL~Her models with different chemical compositions across different wavelengths for the convective parameter sets~A, B, C, and D.}
\label{fig:PLZ}
\end{figure*}

The variation of the theoretical $PL$ and $PW$ relations obtained from the BL~Her models as a function of the different convection parameter sets and metallicity across the different $Gaia$ wavelengths is presented in Fig.~\ref{fig:PLZ}, as well as in Fig.~\ref{fig:PL_diffZ}, and discussed further in the subsequent sections. A comparison of the theoretical $PL$ and $PW$ relations for BL~Her models obtained using linear and non-linear pulsation periods is also presented in Table~\ref{tab:PL_linear}.

\subsection{Effects of convection parameters}

To test for the effects of convection parameters on the theoretical $PL$ and $PW$ relations, we checked whether the slopes of the theoretical relations of the BL~Her models are statistically similar when computed with the different convection parameter sets. We used the standard $t$-test for this, a detailed description of which is provided in \citet{ngeow2015} and briefly summarized below. The assumptions of independent samples with normal distributions and homogeneity of variance hold true for our data sets, and for $PL$ relations in general as shown by \citet{bhardwaj2016a} using quartile-quartile plots and bootstrapping of $PL$ residuals. We also note the advantage of using the $t$-test because of their relative robustness to deviations from assumptions \citep[see e.g.][]{posten1984}.

We defined a $T$ statistic for the comparison of two linear regression slopes, $\hat{W}$ with sample sizes, $n$ and $m$, respectively, as follows:
\begin{equation}
T=\frac{\hat{W}_n-\hat{W}_m}{\sqrt{\mathrm{Var}(\hat{W}_n)+\mathrm{Var}(\hat{W}_m)}},
\label{eq:ttest}
\end{equation}
where $\mathrm{Var}(\hat{W})$ is the variance of the slope. The null hypothesis of equivalent slopes is rejected if $T>t_{\alpha/2,\nu}$ (or the probability of the observed value of the $T$ statistic) is $p<0.05,$ where $t_{\alpha/2,\nu}$ is the critical value under the two-tailed $t$-distribution with 95\% confidence limit ($\alpha$=0.05) and degrees of freedom, $\nu=n+m-4$.

The statistical comparison of the theoretical $PL$ and $PW$ slopes from the BL~Her models with respect to the empirical slopes from $Gaia$ DR3 across the individual $Gaia$ passbands is listed in Table~\ref{tab:PL}. Considering the complete set of BL~Her models, we find that the models with radiative cooling (sets B and D) exhibit statistically similar $PL$ and $PW$ slopes across all the $Gaia$ passbands. This result is similar to what was obtained for the Johnson-Cousin-Glass bands ($BVRIJHKLL'M$) in \citetalias{das2021}. However, this result does not hold true for the subset of low-mass ($0.5-0.6 M_{\odot}$) BL~Her models. The $PL$ slopes for the low-mass BL~Her models are statistically consistent across all the $Gaia$ passbands for the convection parameter sets A, C, and D, but not with the $PL$ slope computed using parameter set B. In addition, we find that only parameter sets A and D exhibit statistically similar $PW$ slopes for the low-mass BL~Her models. 

\begin{table*}
\caption{Comparison of the slopes of the $PL$ and $PW$ relations for BL~Her stars of the mathematical form $M_\lambda=a\log(P)+b$. The theoretical relations are derived for the cases of the complete set of models and for the low mass models. $\varpi>0$ indicates the subset of observed BL~Her stars in the All Sky region with positive parallaxes from $Gaia$ DR3. $N$ is the total number of stars. |$T$| represents the observed value of the $t$-statistic, and $p(t)$ gives the probability of acceptance of the null hypothesis (equal slopes). The bold-faced entries indicate that the null hypothesis of the equivalent $PL$ slopes can be rejected.}
\centering
\scalebox{0.7}{
\begin{tabular}{c c c c c c c c c c c c}
\hline\hline
Band & Source & $a$ & $b$ & $\sigma$ & $N$ & Reference${^\ddagger}$ & Theoretical/ & \multicolumn{4}{c}{(|$T$|, $p(t)$) w.r.t.}\\
& & & & & & & Empirical & Set A & Set B & Set C & Set D\\
\hline \hline
\multicolumn{12}{c}{Complete set of models ($0.5-0.8M_{\odot}$)}\\
\hline
$G$ & $\rm{Z_{all}}$ (Set A) &-1.76$\pm$0.029&-0.263$\pm$0.01&0.262&3266& TW & Theoretical & ... & ... & ... & ...\\
$G$ & $\rm{Z_{all}}$ (Set B) &-1.531$\pm$0.031&-0.261$\pm$0.01&0.237&2260& TW & Theoretical & \textbf{(5.352,0.0)}& ... & ... & ...\\
$G$ & $\rm{Z_{all}}$ (Set C) &-1.63$\pm$0.032&-0.171$\pm$0.011&0.267&2632& TW & Theoretical & \textbf{(2.996,0.003)} & \textbf{(2.231,0.026)} & ... & ...\\
$G$ & $\rm{Z_{all}}$ (Set D) &-1.49$\pm$0.033&-0.183$\pm$0.011&0.255&2122& TW & Theoretical & \textbf{(6.095,0.0)} & (0.895,0.371) & \textbf{(3.049,0.002)} & ...\\
$G$ & All Sky &-0.307$\pm$0.46&15.869$\pm$0.151&1.57&422& Gaia DR3 & Empirical & \textbf{(3.152,0.002)} & \textbf{(2.655,0.008)} & \textbf{(2.869,0.004)} & \textbf{(2.565,0.01)} \\
$G$ & All Sky $(\varpi>0)$&-0.429$\pm$0.464&15.881$\pm$0.151&1.566&415& Gaia DR3 & Empirical & \textbf{(2.863,0.004)} & \textbf{(2.37,0.018)} & \textbf{(2.582,0.01)} & \textbf{(2.281,0.023)} \\
$G$ & LMC &-1.237$\pm$0.151&18.549$\pm$0.043&0.161&61& Gaia DR3 & Empirical & \textbf{(3.401,0.001)} & (1.907,0.057) & \textbf{(2.546,0.011)} & (1.637,0.102)\\
$G$ & SMC &-2.686$\pm$0.721&19.022$\pm$0.214&0.324&16& Gaia DR3 & Empirical & (1.283,0.199) & (1.6,0.11) & (1.463,0.144) & (1.657,0.098)\\
\hline
$G_{BP}$ & $\rm{Z_{all}}$ (Set A) &-1.543$\pm$0.033&-0.077$\pm$0.011&0.293&3266& TW & Theoretical & ... & ... & ... & ...\\
$G_{BP}$ & $\rm{Z_{all}}$ (Set B) &-1.296$\pm$0.034&-0.068$\pm$0.011&0.262&2260& TW & Theoretical & \textbf{(5.193,0.0)}& ... & ... & ...\\
$G_{BP}$ & $\rm{Z_{all}}$ (Set C) &-1.382$\pm$0.036&0.034$\pm$0.012&0.299&2632& TW & Theoretical & \textbf{(3.33,0.001)} & (1.731,0.084) & ... & ...\\
$G_{BP}$ & $\rm{Z_{all}}$ (Set D) &-1.238$\pm$0.037&0.024$\pm$0.013&0.284&2122& TW & Theoretical & \textbf{(6.178,0.0)} & (1.151,0.25) & \textbf{(2.804,0.005)} & ...\\
$G_{BP}$ & All Sky &0.265$\pm$0.593&16.504$\pm$0.195&1.996&404& Gaia DR3 & Empirical & \textbf{(3.044,0.002)} & \textbf{(2.628,0.009)} & \textbf{(2.772,0.006)} & \textbf{(2.53,0.011)}\\
$G_{BP}$ & All Sky $(\varpi>0)$&0.089$\pm$0.596&16.52$\pm$0.194&1.982&397& Gaia DR3 & Empirical & \textbf{(2.734,0.006)} & \textbf{(2.32,0.02)} & \textbf{(2.464,0.014)} & \textbf{(2.222,0.026)}\\
$G_{BP}$ & LMC &-0.633$\pm$0.266&18.648$\pm$0.077&0.258&55& Gaia DR3 & Empirical & \textbf{(3.395,0.001)} & \textbf{(2.472,0.013)} & \textbf{(2.79,0.005)} & \textbf{(2.253,0.024)}\\
$G_{BP}$ & SMC &-2.866$\pm$0.951&19.306$\pm$0.282&0.403&14& Gaia DR3 & Empirical & (1.39,0.165) & (1.65,0.099) & (1.559,0.119) & (1.711,0.087)\\
\hline
$G_{RP}$ & $\rm{Z_{all}}$ (Set A) &-2.015$\pm$0.025&-0.588$\pm$0.009&0.225&3266& TW & Theoretical & ... & ... & ... & ...\\
$G_{RP}$ & $\rm{Z_{all}}$ (Set B) &-1.81$\pm$0.027&-0.594$\pm$0.009&0.209&2260& TW & Theoretical & \textbf{(5.503,0.0)}& ... & ... & ...\\
$G_{RP}$ & $\rm{Z_{all}}$ (Set C) &-1.909$\pm$0.028&-0.523$\pm$0.01&0.231&2632& TW & Theoretical & \textbf{(2.819,0.005)} & \textbf{(2.565,0.01)} & ... & ...\\
$G_{RP}$ & $\rm{Z_{all}}$ (Set D) &-1.779$\pm$0.029&-0.536$\pm$0.01&0.223&2122& TW & Theoretical & \textbf{(6.144,0.0)} & (0.783,0.434) & \textbf{(3.271,0.001)} & ...\\
$G_{RP}$ & All Sky &-0.692$\pm$0.401&15.032$\pm$0.132&1.356&408& Gaia DR3 & Empirical & \textbf{(3.293,0.001)} & \textbf{(2.782,0.005)} & \textbf{(3.028,0.002)} & \textbf{(2.704,0.007)}\\
$G_{RP}$ & All Sky $(\varpi>0)$&-0.801$\pm$0.405&15.043$\pm$0.132&1.354&401& Gaia DR3 & Empirical & \textbf{(2.992,0.003)} & \textbf{(2.486,0.013)} & \textbf{(2.729,0.006)} & \textbf{(2.409,0.016)}\\
$G_{RP}$ & LMC &-1.769$\pm$0.205&18.149$\pm$0.061&0.206&55& Gaia DR3 & Empirical & (1.191,0.234) & (0.198,0.843) & (0.677,0.499) & (0.048,0.961)\\
$G_{RP}$ & SMC &-2.777$\pm$0.701&18.602$\pm$0.208&0.297&14& Gaia DR3 & Empirical & (1.086,0.277) & (1.378,0.168) & (1.237,0.216) & (1.422,0.155)\\
\hline
$W(G,G_{BP}-G_{RP})$ & $\rm{Z_{all}}$ (Set A) &-2.656$\pm$0.018&-1.234$\pm$0.006&0.159&3266& TW & Theoretical & ... & ... & ... & ...\\
$W(G,G_{BP}-G_{RP})$ & $\rm{Z_{all}}$ (Set B) &-2.507$\pm$0.021&-1.261$\pm$0.007&0.159&2260& TW & Theoretical & \textbf{(5.432,0.0)}& ... & ... & ...\\
$W(G,G_{BP}-G_{RP})$ & $\rm{Z_{all}}$ (Set C) &-2.633$\pm$0.019&-1.231$\pm$0.007&0.163&2632& TW & Theoretical & (0.887,0.375) & \textbf{(4.412,0.0)} & ... & ...\\
$W(G,G_{BP}-G_{RP})$ & $\rm{Z_{all}}$ (Set D) &-2.517$\pm$0.021&-1.247$\pm$0.007&0.165&2122& TW & Theoretical & \textbf{(4.98,0.0)} & (0.341,0.733) & \textbf{(3.994,0.0)} & ...\\
$W(G,G_{BP}-G_{RP})$ & All Sky &-2.031$\pm$0.3&12.932$\pm$0.098&0.982&409& Gaia DR3 & Empirical & \textbf{(2.08,0.038)} & (1.583,0.114) & \textbf{(2.003,0.045)} & (1.616,0.106)\\
$W(G,G_{BP}-G_{RP})$ & All Sky$(\varpi>0)$ &-2.044$\pm$0.304&12.935$\pm$0.098&0.981&402& Gaia DR3 & Empirical & \textbf{(2.01,0.045)} & (1.519,0.129) & (1.934,0.053) & (1.552,0.121)\\
$W(G,G_{BP}-G_{RP})$ & LMC &-2.398$\pm$0.146&17.379$\pm$0.044&0.144&58& Gaia DR3 & Empirical & (1.754,0.08) & (0.739,0.46) & (1.596,0.111) & (0.807,0.42)\\
$W(G,G_{BP}-G_{RP})$ & SMC &-2.456$\pm$0.439&17.662$\pm$0.13&0.186&14& Gaia DR3 & Empirical & (0.455,0.649) & (0.116,0.908) & (0.403,0.687) & (0.139,0.89)\\
\hline
\multicolumn{12}{c}{Low-mass models only ($0.5-0.6M_{\odot}$)}\\
\hline

$G$ & $\rm{Z_{all}}$ (Set A) &-1.346$\pm$0.045&-0.192$\pm$0.015&0.222&1050& TW & Theoretical & ... & ... & ... & ...\\
$G$ & $\rm{Z_{all}}$ (Set B) &-1.223$\pm$0.04&-0.151$\pm$0.013&0.178&707& TW & Theoretical & \textbf{(2.049,0.041)}& ... & ... & ...\\
$G$ & $\rm{Z_{all}}$ (Set C) &-1.465$\pm$0.048&-0.014$\pm$0.016&0.234&856& TW & Theoretical & (1.829,0.068) & \textbf{(3.894,0.0)} & ... & ...\\
$G$ & $\rm{Z_{all}}$ (Set D) &-1.432$\pm$0.046&-0.007$\pm$0.016&0.222&711& TW & Theoretical & (1.343,0.179) & \textbf{(3.42,0.001)} & (0.498,0.619) & ...\\
$G$ & All Sky &-0.307$\pm$0.46&15.869$\pm$0.151&1.57&422& Gaia DR3 & Empirical & \textbf{(2.248,0.025)} & \textbf{(1.984,0.048)} & \textbf{(2.504,0.012)} & \textbf{(2.434,0.015)}\\
$G$ & All Sky $(\varpi>0)$&-0.429$\pm$0.464&15.881$\pm$0.151&1.566&415& Gaia DR3 & Empirical & \textbf{(1.967,0.049)} & (1.705,0.088) & \textbf{(2.221,0.027)} & (2.151,0.032)\\
$G$ & LMC &-1.237$\pm$0.151&18.549$\pm$0.043&0.161&61& Gaia DR3 & Empirical & (0.692,0.489) & (0.09,0.929) & (1.439,0.15) & (1.235,0.217)\\
$G$ & SMC &-2.686$\pm$0.721&19.022$\pm$0.214&0.324&16& Gaia DR3 & Empirical & (1.855,0.064) & \textbf{(2.026,0.043)} & (1.69,0.091) & (1.736,0.083)\\
\hline
$G_{BP}$ & $\rm{Z_{all}}$ (Set A) &-1.074$\pm$0.051&-0.017$\pm$0.017&0.254&1050& TW & Theoretical & ... & ... & ... & ...\\
$G_{BP}$ & $\rm{Z_{all}}$ (Set B) &-0.952$\pm$0.045&0.037$\pm$0.015&0.202&707& TW & Theoretical & (1.78,0.075)& ... & ... & ...\\
$G_{BP}$ & $\rm{Z_{all}}$ (Set C) &-1.197$\pm$0.055&0.194$\pm$0.019&0.269&856& TW & Theoretical & (1.641,0.101) & \textbf{(3.438,0.001)} & ... & ...\\
$G_{BP}$ & $\rm{Z_{all}}$ (Set D) &-1.179$\pm$0.053&0.206$\pm$0.018&0.254&711& TW & Theoretical & (1.426,0.154) & \textbf{(3.246,0.001)} & (0.234,0.815) & ...\\
$G_{BP}$ & All Sky &0.265$\pm$0.593&16.504$\pm$0.195&1.996&404& Gaia DR3 & Empirical & \textbf{(2.25,0.025)} & \textbf{(2.046,0.041)} & \textbf{(2.455,0.014)} & \textbf{(2.425,0.015)}\\
$G_{BP}$ & All Sky $(\varpi>0)$&0.089$\pm$0.596&16.52$\pm$0.194&1.982&397& Gaia DR3 & Empirical & (1.944,0.052) & (1.742,0.082) & \textbf{(2.149,0.032)} &  \textbf{(2.119,0.034)}\\
$G_{BP}$ & LMC &-0.633$\pm$0.266&18.648$\pm$0.077&0.258&55& Gaia DR3 & Empirical & (1.628,0.104) & (1.182,0.237) &  \textbf{(2.076,0.038)} &  \textbf{(2.013,0.044)}\\
$G_{BP}$  & SMC &-2.866$\pm$0.951&19.306$\pm$0.282&0.403&14& Gaia DR3 & Empirical & (1.882,0.06) &  \textbf{(2.01,0.045)} & (1.752,0.08) & (1.771,0.077)\\
\hline
$G_{RP}$ & $\rm{Z_{all}}$ (Set A) &-1.669$\pm$0.037&-0.502$\pm$0.012&0.182&1050& TW & Theoretical & ... & ... & ... & ...\\
$G_{RP}$ & $\rm{Z_{all}}$ (Set B) &-1.549$\pm$0.033&-0.475$\pm$0.011&0.149&707& TW & Theoretical & \textbf{(2.405,0.016)}& ... & ... & ...\\
$G_{RP}$ & $\rm{Z_{all}}$ (Set C) &-1.768$\pm$0.039&-0.367$\pm$0.013&0.193&856& TW & Theoretical & (1.845,0.065) & \textbf{(4.237,0.0)} & ... & ...\\
$G_{RP}$ & $\rm{Z_{all}}$ (Set D) &-1.725$\pm$0.038&-0.364$\pm$0.013&0.183&711& TW & Theoretical & (1.057,0.29) & \textbf{(3.451,0.001)} & (0.786,0.432) & ...\\
$G_{RP}$ & All Sky &-0.692$\pm$0.401&15.032$\pm$0.132&1.356&408& Gaia DR3 & Empirical & \textbf{(2.426,0.015)} & \textbf{(2.13,0.033)} & \textbf{(2.671,0.008)} & \textbf{(2.565,0.01)}\\
$G_{RP}$ & All Sky $(\varpi>0)$&-0.801$\pm$0.405&15.043$\pm$0.132&1.354&401& Gaia DR3 & Empirical & \textbf{(2.134,0.033)} & (1.841,0.066) & \textbf{(2.377,0.018)} & \textbf{(2.272,0.023)}\\
$G_{RP}$ & LMC &-1.769$\pm$0.205&18.149$\pm$0.061&0.206&55& Gaia DR3 & Empirical & (0.48,0.631) & (1.06,0.29) & (0.005,0.996) & (0.211,0.833)\\
$G_{RP}$ & SMC &-2.777$\pm$0.701&18.602$\pm$0.208&0.297&14& Gaia DR3 & Empirical & (1.578,0.115) & (1.75,0.081) & (1.437,0.151) & (1.499,0.134)\\
\hline
$W(G,G_{BP}-G_{RP})$ & $\rm{Z_{all}}$ (Set A) &-2.476$\pm$0.022&-1.113$\pm$0.007&0.107&1050& TW & Theoretical & ... & ... & ... & ...\\
$W(G,G_{BP}-G_{RP})$ & $\rm{Z_{all}}$ (Set B) &-2.358$\pm$0.022&-1.124$\pm$0.007&0.1&707& TW & Theoretical & \textbf{(3.799,0.0)}& ... & ... & ...\\
$W(G,G_{BP}-G_{RP})$ & $\rm{Z_{all}}$ (Set C) &-2.551$\pm$0.023&-1.08$\pm$0.008&0.113&856& TW & Theoretical & \textbf{(2.37,0.018)} & \textbf{(6.011,0.0)} & ... & ...\\
$W(G,G_{BP}-G_{RP})$ & $\rm{Z_{all}}$ (Set D) &-2.47$\pm$0.023&-1.089$\pm$0.008&0.112&711& TW & Theoretical & (0.207,0.836) & \textbf{(3.448,0.001)} & \textbf{(2.481,0.013)} & ...\\
$W(G,G_{BP}-G_{RP})$ & All Sky &-2.031$\pm$0.3&12.932$\pm$0.098&0.982&409& Gaia DR3 & Empirical & (1.479,0.139) & (1.087,0.277) & (1.728,0.084) & (1.459,0.145)\\
$W(G,G_{BP}-G_{RP})$ & All Sky $(\varpi>0)$&-2.044$\pm$0.304&12.935$\pm$0.098&0.981&402& Gaia DR3 & Empirical & (1.417,0.157) & (1.03,0.303) & (1.663,0.097) & (1.397,0.163)\\
$W(G,G_{BP}-G_{RP})$ & LMC &-2.398$\pm$0.146&17.379$\pm$0.044&0.144&58& Gaia DR3 & Empirical & (0.528,0.597) & (0.271,0.787) & (1.035,0.301) & (0.487,0.626)\\
$W(G,G_{BP}-G_{RP})$ & SMC &-2.456$\pm$0.439&17.662$\pm$0.13&0.186&14& Gaia DR3 & Empirical & (0.046,0.964) & (0.223,0.824) & (0.216,0.829) & (0.032,0.975)\\
\hline
\end{tabular}}
\label{tab:PL}
\end{table*}

\begin{figure*}
\centering
\includegraphics[scale = 1]{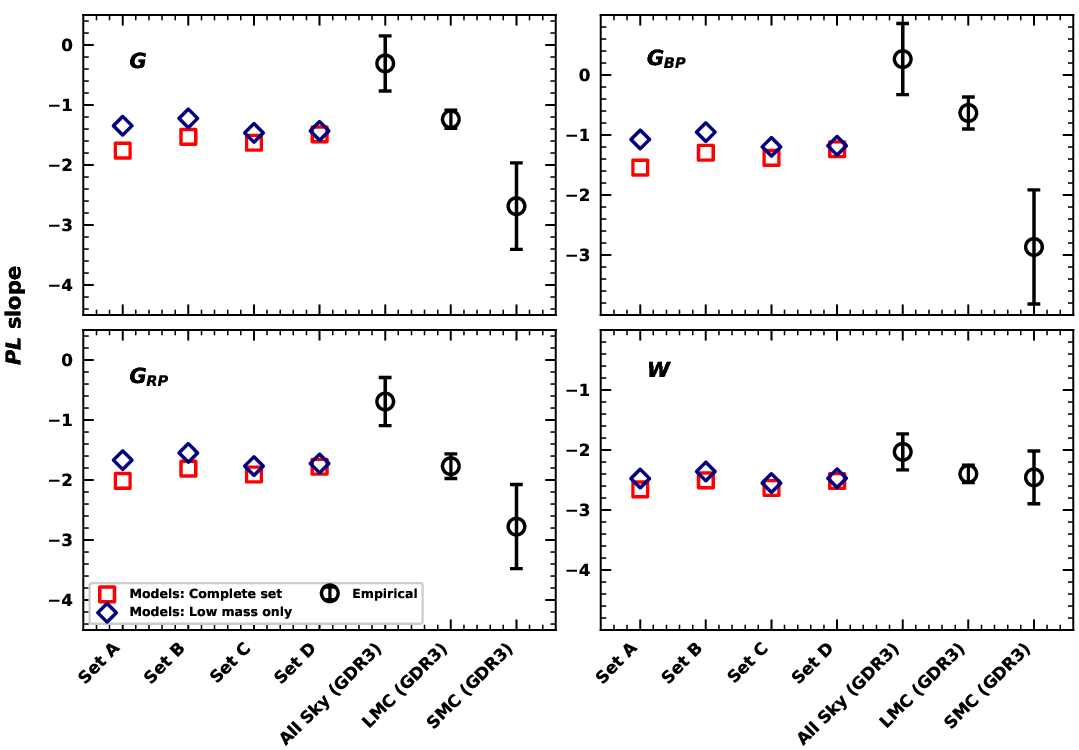}
\caption{Comparison of the theoretical $PL$ and $PW$ slopes of the BL~Her models computed using the complete range of metallicities ($-2.0\; \mathrm{dex} \leq \mathrm{[Fe/H]} \leq 0.0\; \mathrm{dex}$), with the empirical slopes spread across the different $Gaia$ wavelengths. The y-scale is same (5 units) in each panel for a relative comparison. Note: the empirical $PL$ slopes in the $Gaia$ passbands are obtained using apparent magnitudes not corrected for extinction.}
\label{fig:PL_slopes}
\end{figure*}

A comparison of the theoretical $PL$ and $PW$ slopes of the BL~Her models with the empirical slopes across the different $Gaia$ wavelengths is displayed in Fig.~\ref{fig:PL_slopes}.  We have also considered a subset of observed BL~Her stars with positive parallaxes in the All Sky region for a comparison of the theoretical slopes with the empirical slopes. From Table~\ref{tab:PL} and Fig.~\ref{fig:PL_slopes}, we find that the BL~Her stars in the All Sky region exhibit statistically different $PL$ slopes compared to the theoretical $PL$ slopes computed using the four sets of convection parameters. This is also true for the theoretical slopes obtained using the subset of low-mass models only, with the exception of the stars with positive parallaxes exhibiting statistically similar $PL$ slopes with theoretical slopes computed using the convection parameter set B. The statistically different empirical $PL$ slopes of the BL~Her stars in the All Sky region is expected because the empirical $PL$ slopes in the $Gaia$ passbands are obtained using apparent magnitudes not corrected for extinction. However, the empirical $PW$ relations from the BL~Her stars in the All Sky region mostly exhibit statistically similar slopes with the theoretical $PW$ slopes computed using the four sets of convection parameters, with a few exceptions. 

The empirical $PL$ and $PW$ slopes from BL~Her stars in the LMC and the SMC mostly agree well with the theoretical relations computed using the different convection parameter sets across all the $Gaia$ passbands. BL~Her stars in the Magellanic Clouds exhibiting statistically similar $PL$ slopes with those from the models in the $Gaia$ passbands (despite not being corrected for extinction) can be explained by the presence of small average foreground reddening values in the LMC and the SMC which affects the $PL$ zero points but keeps the $PL$ slopes unaffected \citep{ripepi2019}. The fact that the empirical $PL$ and $PW$ slopes from the BL~Her stars are statistically consistent with the theoretical slopes across the four different sets of convection parameters, while the theoretical slopes from the different convection sets themselves exhibit statistical differences amongst each other, indicates that the uncertainties from observations dominate over the uncertainty on the theoretical fits. This is especially true for the case of BL~Her stars in the SMC where the empirical relations exhibit large uncertainties. However, we note here that the large dispersion in the $PL$ and $PW$ relations from the BL~Her stars in the SMC arises because of the large depth along its line of sight \citep[see][and references therein]{ripepi2022}. Larger data sets of BL~Her stars with more precise observations would prove useful in the future for constraining the stellar pulsation codes better with respect to the choice of convection parameters and in the subsequent estimation of the best-fit model parameters.

The $Gaia$ passbands are, in increasing order of the central effective wavelengths ($\lambda_{\rm eff}$) of the respective passbands, as follows: $G_{BP} < G < G_{RP}$ \citep{rodrigo2012, rodrigo2020}. From Table~\ref{tab:PL}, we find that the BL~Her models exhibit steeper theoretical $PL$ slopes in the $Gaia$ passbands with increasing wavelengths. This result is consistent with the theoretical relations for the Johnson-Cousin-Glass bands ($UBVRIJHKLL'M$) in \citetalias{das2021} and also in support of similar empirical relations obtained for RR~Lyraes as shown in \citet{beaton2018b, neeley2017, bhardwaj2020a}. In addition, we also observe a decrease in the dispersion in the theoretical $PL$ relations for BL~Her models with increasing wavelength, with the smallest dispersion for the theoretical $PW$ relations. As noted in \citet{catelan2004, madore2012, marconi2015}, the smaller dispersion in the $PL$ relations as a function of increasing wavelength is caused by the decrease of the width in the instability strip itself at longer wavelengths.

\begin{table}[ht!]
\caption{$PLZ$ relations for BL~Her models of the mathematical form $M_\lambda=\alpha+\beta\log(P)+\gamma\mathrm{[Fe/H]}$ for different wavelengths using different convective parameter sets.}
\centering
\scalebox{0.8}{
\begin{tabular}{c c c c c c}
\hline\hline
Band & $\alpha$ & $\beta$ & $\gamma$ & $\sigma$ & $N$\\
\hline \hline
\multicolumn{6}{c}{Complete set of models ($0.5-0.8M_{\odot}$)}\\
\hline
\multicolumn{6}{c}{Convection set A}\\
\hline
$G$ & -0.262 $\pm$ 0.012 & -1.761 $\pm$ 0.03 & 0.001 $\pm$ 0.007 & 0.262 & 3266\\
$G_{BP}$ & -0.058 $\pm$ 0.014 & -1.555 $\pm$ 0.033 & 0.018 $\pm$ 0.008 & 0.292 & 3266\\
$G_{RP}$ & -0.601 $\pm$ 0.011 & -2.007 $\pm$ 0.026 & -0.012 $\pm$ 0.006 & 0.225 & 3266\\
$W$ & -1.294 $\pm$ 0.007 & -2.619 $\pm$ 0.018 & -0.056 $\pm$ 0.004 & 0.154 & 3266\\
\hline
\multicolumn{6}{c}{Convection set B}\\
\hline
$G$ & -0.27 $\pm$ 0.013 & -1.527 $\pm$ 0.031 & -0.008 $\pm$ 0.007 & 0.237 & 2260\\
$G_{BP}$ & -0.06 $\pm$ 0.014 & -1.3 $\pm$ 0.035 & 0.008 $\pm$ 0.008 & 0.262 & 2260\\
$G_{RP}$ & -0.615 $\pm$ 0.011 & -1.8 $\pm$ 0.027 & -0.021 $\pm$ 0.006 & 0.208 & 2260\\
$W$ & -1.325 $\pm$ 0.008 & -2.477 $\pm$ 0.02 & -0.063 $\pm$ 0.005 & 0.153 & 2260\\
\hline
\multicolumn{6}{c}{Convection set C}\\
\hline
$G$ & -0.174 $\pm$ 0.014 & -1.628 $\pm$ 0.033 & -0.003 $\pm$ 0.008 & 0.267 & 2632\\
$G_{BP}$ & 0.051 $\pm$ 0.015 & -1.394 $\pm$ 0.036 & 0.015 $\pm$ 0.009 & 0.299 & 2632\\
$G_{RP}$ & -0.541 $\pm$ 0.012 & -1.896 $\pm$ 0.028 & -0.016 $\pm$ 0.007 & 0.231 & 2632\\
$W$ & -1.298 $\pm$ 0.008 & -2.581 $\pm$ 0.019 & -0.063 $\pm$ 0.005 & 0.158 & 2632\\
\hline
\multicolumn{6}{c}{Convection set D}\\
\hline
$G$ & -0.188 $\pm$ 0.014 & -1.488 $\pm$ 0.033 & -0.004 $\pm$ 0.008 & 0.255 & 2122\\
$G_{BP}$ & 0.038 $\pm$ 0.016 & -1.246 $\pm$ 0.037 & 0.014 $\pm$ 0.009 & 0.284 & 2122\\
$G_{RP}$ & -0.554 $\pm$ 0.012 & -1.769 $\pm$ 0.029 & -0.017 $\pm$ 0.007 & 0.223 & 2122\\
$W$ & -1.312 $\pm$ 0.009 & -2.481 $\pm$ 0.021 & -0.063 $\pm$ 0.005 & 0.16 & 2122\\
\hline
\multicolumn{6}{c}{Low-mass models only ($0.5-0.6M_{\odot}$)}\\
\hline
\multicolumn{6}{c}{Convection set A}\\
\hline
$G$ & -0.21 $\pm$ 0.018 & -1.333 $\pm$ 0.045 & -0.016 $\pm$ 0.01 & 0.222 & 1050 \\
$G_{BP}$ & -0.018 $\pm$ 0.021 & -1.073 $\pm$ 0.052 & -0.001 $\pm$ 0.012 & 0.254 & 1050 \\
$G_{RP}$ & -0.532 $\pm$ 0.015 & -1.647 $\pm$ 0.037 & -0.028 $\pm$ 0.008 & 0.181 & 1050 \\
$W$ & -1.187 $\pm$ 0.008 & -2.423 $\pm$ 0.02 & -0.068 $\pm$ 0.004 & 0.097 & 1050\\
\hline
\multicolumn{6}{c}{Convection set B}\\
\hline
$G$ & -0.164 $\pm$ 0.016 & -1.218 $\pm$ 0.04 & -0.014 $\pm$ 0.01 & 0.178 & 707\\
$G_{BP}$ & 0.04 $\pm$ 0.018 & -0.953 $\pm$ 0.046 & 0.003 $\pm$ 0.011 & 0.202 & 707\\
$G_{RP}$ & -0.501 $\pm$ 0.013 & -1.54 $\pm$ 0.033 & -0.026 $\pm$ 0.008 & 0.148 & 707\\
$W$ & -1.191 $\pm$ 0.008 & -2.333 $\pm$ 0.02 & -0.07 $\pm$ 0.005 & 0.088 & 707\\
\hline
\multicolumn{6}{c}{Convection set C}\\
\hline
$G$ & -0.022 $\pm$ 0.02 & -1.459 $\pm$ 0.049 & -0.008 $\pm$ 0.012 & 0.234 & 856\\
$G_{BP}$ & 0.206 $\pm$ 0.023 & -1.206 $\pm$ 0.056 & 0.012 $\pm$ 0.014 & 0.268 & 856\\
$G_{RP}$ & -0.39 $\pm$ 0.017 & -1.751 $\pm$ 0.04 & -0.023 $\pm$ 0.01 & 0.192 & 856\\
$W$ & -1.155 $\pm$ 0.009 & -2.495 $\pm$ 0.021 & -0.073 $\pm$ 0.005 & 0.101 & 856\\
\hline
\multicolumn{6}{c}{Convection set D}\\
\hline
$G$ & -0.011 $\pm$ 0.02 & -1.43 $\pm$ 0.047 & -0.005 $\pm$ 0.012 & 0.222 & 711\\
$G_{BP}$ & 0.219 $\pm$ 0.023 & -1.184 $\pm$ 0.053 & 0.015 $\pm$ 0.014 & 0.254 & 711\\
$G_{RP}$ & -0.382 $\pm$ 0.016 & -1.718 $\pm$ 0.038 & -0.02 $\pm$ 0.01 & 0.183 & 711\\
$W$ & -1.154 $\pm$ 0.009 & -2.447 $\pm$ 0.021 & -0.071 $\pm$ 0.006 & 0.101 & 711\\
\hline
\end{tabular}}
\label{tab:PLZ}
\end{table}

\begin{table}[ht!]
\caption{$PLZ$ relations for BL~Her models of the mathematical form $M_\lambda=\alpha+\beta\log(P)+\gamma\mathrm{[Fe/H]}$ in the low- and the high-metallicity regimes for different wavelengths using different convective parameter sets.}
\centering
\scalebox{0.8}{
\begin{tabular}{c c c c c c}
\hline\hline
Band & $\alpha$ & $\beta$ & $\gamma$ & $\sigma$ & $N$\\
\hline \hline
\multicolumn{6}{c}{Low-metallicity regime ($Z=0.00135, 0.00061, 0.00043, 0.00014$)}\\
\hline
\multicolumn{6}{c}{Convection set A}\\
\hline
$G$ & -0.288 $\pm$ 0.029 & -1.749 $\pm$ 0.042 & -0.014 $\pm$ 0.018 & 0.263 & 1734\\
$G_{BP}$ & -0.094 $\pm$ 0.032 & -1.558 $\pm$ 0.047 & -0.007 $\pm$ 0.02 & 0.292 & 1734\\
$G_{RP}$ & -0.616 $\pm$ 0.025 & -1.985 $\pm$ 0.036 & -0.018 $\pm$ 0.015 & 0.227 & 1734\\
$W$ & -1.278 $\pm$ 0.017 & -2.561 $\pm$ 0.025 & -0.035 $\pm$ 0.01 & 0.157 & 1734\\
\hline
\multicolumn{6}{c}{Convection set B}\\
\hline
$G$ & -0.271 $\pm$ 0.03 & -1.526 $\pm$ 0.041 & -0.009 $\pm$ 0.018 & 0.229 & 1217\\
$G_{BP}$ & -0.069 $\pm$ 0.032 & -1.319 $\pm$ 0.045 & -0.001 $\pm$ 0.02 & 0.25 & 1217\\
$G_{RP}$ & -0.609 $\pm$ 0.026 & -1.785 $\pm$ 0.036 & -0.013 $\pm$ 0.016 & 0.202 & 1217\\
$W$ & -1.296 $\pm$ 0.02 & -2.412 $\pm$ 0.027 & -0.032 $\pm$ 0.012 & 0.153 & 1217\\
\hline
\multicolumn{6}{c}{Convection set C}\\
\hline
$G$ & -0.218 $\pm$ 0.032 & -1.52 $\pm$ 0.045 & -0.012 $\pm$ 0.02 & 0.254 & 1292\\
$G_{BP}$ & -0.007 $\pm$ 0.036 & -1.287 $\pm$ 0.05 & -0.003 $\pm$ 0.022 & 0.281 & 1292\\
$G_{RP}$ & -0.571 $\pm$ 0.028 & -1.798 $\pm$ 0.039 & -0.018 $\pm$ 0.017 & 0.222 & 1292\\
$W$ & -1.29 $\pm$ 0.02 & -2.491 $\pm$ 0.028 & -0.041 $\pm$ 0.012 & 0.158 & 1292\\
\hline
\multicolumn{6}{c}{Convection set D}\\
\hline
$G$ & -0.217 $\pm$ 0.032 & -1.438 $\pm$ 0.042 & -0.013 $\pm$ 0.02 & 0.235 & 1094\\
$G_{BP}$ & -0.003 $\pm$ 0.035 & -1.212 $\pm$ 0.046 & -0.006 $\pm$ 0.022 & 0.256 & 1094\\
$G_{RP}$ & -0.57 $\pm$ 0.029 & -1.714 $\pm$ 0.037 & -0.017 $\pm$ 0.018 & 0.209 & 1094\\
$W$ & -1.293 $\pm$ 0.022 & -2.392 $\pm$ 0.029 & -0.034 $\pm$ 0.014 & 0.161 & 1094\\
\hline
\multicolumn{6}{c}{High-metallicity regime ($Z=0.01300, 0.00834, 0.00424$)}\\
\hline
\multicolumn{6}{c}{Convection set A}\\
\hline
$G$ & -0.247 $\pm$ 0.018 & -1.779 $\pm$ 0.042 & 0.036 $\pm$ 0.033 & 0.26 & 1532\\
$G_{BP}$ & -0.043 $\pm$ 0.02 & -1.561 $\pm$ 0.047 & 0.068 $\pm$ 0.037 & 0.292 & 1532\\
$G_{RP}$ & -0.588 $\pm$ 0.015 & -2.031 $\pm$ 0.036 & 0.009 $\pm$ 0.028 & 0.222 & 1532\\
$W$ & -1.283 $\pm$ 0.01 & -2.672 $\pm$ 0.024 & -0.077 $\pm$ 0.019 & 0.15 & 1532\\
\hline
\multicolumn{6}{c}{Convection set B}\\
\hline
$G$ & -0.272 $\pm$ 0.019 & -1.527 $\pm$ 0.048 & -0.023 $\pm$ 0.037 & 0.247 & 1043\\
$G_{BP}$ & -0.066 $\pm$ 0.021 & -1.278 $\pm$ 0.054 & 0.004 $\pm$ 0.042 & 0.275 & 1043\\
$G_{RP}$ & -0.615 $\pm$ 0.016 & -1.816 $\pm$ 0.042 & -0.042 $\pm$ 0.033 & 0.214 & 1043\\
$W$ & -1.316 $\pm$ 0.012 & -2.549 $\pm$ 0.03 & -0.11 $\pm$ 0.023 & 0.152 & 1043\\
\hline
\multicolumn{6}{c}{Convection set C}\\
\hline
$G$ & -0.139 $\pm$ 0.02 & -1.727 $\pm$ 0.047 & 0.005 $\pm$ 0.038 & 0.278 & 1340\\
$G_{BP}$ & 0.09 $\pm$ 0.022 & -1.494 $\pm$ 0.053 & 0.036 $\pm$ 0.043 & 0.314 & 1340\\
$G_{RP}$ & -0.512 $\pm$ 0.017 & -1.983 $\pm$ 0.04 & -0.019 $\pm$ 0.032 & 0.238 & 1340\\
$W$ & -1.282 $\pm$ 0.011 & -2.655 $\pm$ 0.026 & -0.099 $\pm$ 0.021 & 0.155 & 1340\\
\hline
\multicolumn{6}{c}{Convection set D}\\
\hline
$G$ & -0.175 $\pm$ 0.022 & -1.545 $\pm$ 0.053 & -0.042 $\pm$ 0.043 & 0.275 & 1028\\
$G_{BP}$& 0.048 $\pm$ 0.024 & -1.288 $\pm$ 0.06 & -0.018 $\pm$ 0.048 & 0.311 & 1028\\
$G_{RP}$ & -0.541 $\pm$ 0.019 & -1.832 $\pm$ 0.046 & -0.058 $\pm$ 0.036 & 0.236 & 1028\\
$W$ & -1.293 $\pm$ 0.012 & -2.579 $\pm$ 0.03 & -0.118 $\pm$ 0.024 & 0.156 & 1028\\
\hline
\end{tabular}}
\label{tab:PLZ_metallicityregime}
\end{table}

\subsection{Effect of metallicity}

To test for the effect of metallicity on the $PL$ and $PW$ relations, we derive $PLZ$ and $PWZ$ relations for the BL~Her models of the mathematical form:
\begin{equation}
\begin{aligned}
M_\lambda =& \alpha+\beta\log(P)+\gamma\mathrm{[Fe/H]}\quad (\textrm{for} \; \lambda=G,G_{BP},G_{RP}); \\
W =& \alpha+\beta\log(P)+\gamma\mathrm{[Fe/H]}.\\
\end{aligned}
\end{equation}
These results are summarised in Table~\ref{tab:PLZ}. In both the cases for the complete set of models and for the low-mass models only, we find the metallicity contribution to the $PLZ$ relations to be negligible within 3$\sigma$ uncertainties, except for a few cases using $G_{RP}$. However, we always find a small but significant negative coefficient of metallicity in the $PWZ$ relations for the BL~Her models using the four sets of convection parameters and for both the cases of the complete set of models and the low-mass models.

The relatively larger metallicity dependence in the $PWZ$ relations from the BL~Her models using the $Gaia$ passbands is very interesting. This is because a similar result is also exhibited by the $PW_{Gaia}Z$ relations from the classical Cepheids \citep[see][]{ripepi2022, breuval2022, trentin2024}.

We investigate why a small but significant metallicity contribution in the Wesenheit relations occurs when there is no significant effect of metallicity in the individual $Gaia$ passbands. To this end, we separated the models in the low-metallicity regime ($Z=0.00135, 0.00061, 0.00043, 0.00014$)\footnote{\label{note1} For equivalent [Fe/H] range, see Table~\ref{tab:composition}} and the high-metallicity regime ($Z=0.01300, 0.00834, 0.00424$)\footref{note1} and obtained their $PLZ$ relations, the results of which are listed in Table~\ref{tab:PLZ_metallicityregime}. The metallicity coefficients of the $PW$ relations using the $Gaia$ passbands are mostly consistent with zero within $3\sigma$ uncertainties for the BL~Her models in the low-metallicity regime, but clearly exhibit a small but significant and negative metallicity coefficient for the BL~Her models in the high-metallicity regime, thereby contributing to an overall small metallicity effect when considering the complete set of BL~Her models. A possible explanation for the difference in the behaviour of the metallicity contributions in the different metallicity regimes could be because of the increased sensitivity of bolometric corrections to metallicities at wavelengths shorter than the $V$ band \citep{gray2005, kudritzki2008}. Although beyond the scope of the present study, it would be important to investigate the effect of the different model atmospheres and their subsequent impact on the transformations of the bolometric light curves in the future.

\begin{figure*}
\centering
\includegraphics[scale = 0.95]{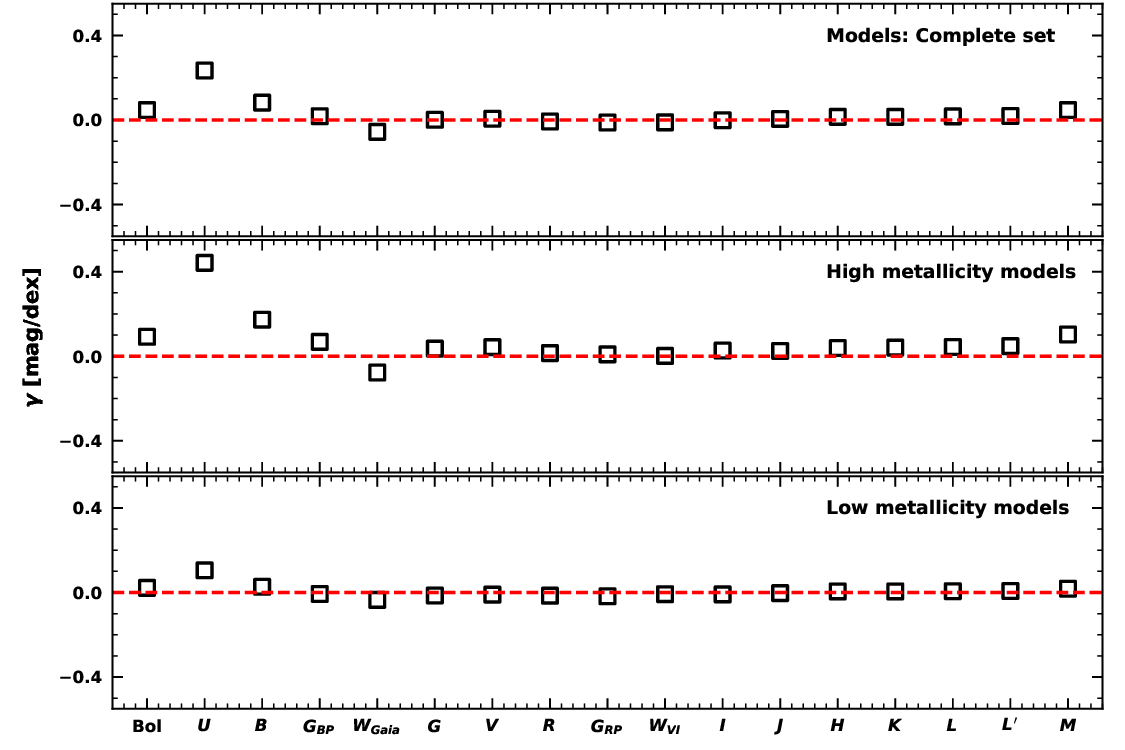}
\caption{Contribution of the $\gamma$ term (metallicity effect) obtained from the theoretical $PLZ$ relations for BL~Her models computed using the convection parameter set~A across the bolometric (Bol), Johnson-Cousins-Glass ($UBVRIJHKLL'M$) and $Gaia$ ($GG_{BP}G_{RP}$) passbands. $W_{Gaia}$ represents the $Gaia$ Wesenheit magnitude $W(G, G_{BP} - G_{RP})$. The different panels present results from the complete set of models compared with the high-metallicity models ($Z=0.00424, 0.00834, 0.01300$) and the low-metallicity models ($Z=0.00014, 0.00043, 0.00061,  0.00135$). The x-axis is in increasing order of the central effective wavelengths ($\lambda_{\rm eff}$) of the respective passbands as provided by the SVO Filter Profile Service \citep{rodrigo2012, rodrigo2020}.}
\label{fig:gamma_metallicity}
\end{figure*}

The contribution of the $\gamma$ term (metallicity effect) obtained from the theoretical $PLZ$ relations for BL~Her models computed using the convection parameter set~A is displayed in Fig.~\ref{fig:gamma_metallicity} as a function of wavelength. The $\gamma$ terms for the $PLZ/PWZ$ relations in the $Gaia$ passbands are obtained from this present analysis. In addition,  we use the $\gamma$ terms for the $PLZ$ relations in the bolometric and the Johnson-Cousins-Glass ($UBVRIJHKLL'M$) bands from \citetalias{das2021}. $W_{Gaia}$ represents the $Gaia$ Wesenheit magnitude $W(G, G_{BP} - G_{RP})=G - 1.90(G_{BP}-G_{RP})$ \citep{ripepi2019} while $W_{VI} = I - 1.55(V-I)$ \citep{inno2013}. From Fig.~\ref{fig:gamma_metallicity}, we find that the metallicity coefficient in the high-metallicity regime is indeed much stronger as compared to the low-metallicity regime at wavelengths shorter than $V$ band, with the strongest effect at $U$ and $B$ bands.

\section{Mass-luminosity relations}
\label{sec:ml}

We explore the dependence of the mass-luminosity ($ML$) relation on metallicity in our BL~Her models, following the work of \citet{bono2020}, where they found a marginal effect of metallicity on the $ML$ relation of T2Cs. To test for the effect of metallicity on the $ML$ relations, we derive $MLZ$ relations in the form:
\begin{equation}
W=\alpha' + \beta' \log(M/M_{\odot})+\gamma' \mathrm{[Fe/H]}
,\end{equation}
for the four different convection parameter sets, where $W$ is the Wesenheit magnitude using the $Gaia$ passbands, as defined in Eq.~\ref{eq:wesenheit}. We find the following relations for set A:
\begin{equation}
\begin{aligned}
W={} & -(2.614 \pm 0.022)-(2.487 \pm 0.107)\log(M/M_{\odot})\\
&-(0.150 \pm 0.010)\mathrm{[Fe/H]} \quad (N=3266; \sigma=0.399),
\end{aligned}
\label{eq:MLZ-A}
\end{equation}
for set B:
\begin{equation}
\begin{aligned}
W={} & -(2.492 \pm 0.027)-(2.283 \pm 0.129)\log(M/M_{\odot})\\
&-(0.131 \pm 0.012)\mathrm{[Fe/H]} \quad (N=2260; \sigma=0.397),
\end{aligned}
\end{equation}
for set C:
\begin{equation}
\begin{aligned}
W={} & -(2.614 \pm 0.026)-(2.395 \pm 0.124)\log(M/M_{\odot})\\
&-(0.191 \pm 0.012)\mathrm{[Fe/H]} \quad (N=2632; \sigma=0.414),
\end{aligned}
\end{equation}
and for set D:
\begin{equation}
\begin{aligned}
W={} & -(2.511 \pm 0.029)-(2.141 \pm 0.140)\log(M/M_{\odot})\\
&-(0.156 \pm 0.013)\mathrm{[Fe/H]} \quad (N=2122; \sigma=0.418).
\end{aligned}
\label{eq:MLZ-D}
\end{equation}

The coefficients of the metallicity term from Eqs.~\ref{eq:MLZ-A}-\ref{eq:MLZ-D} indeed suggest a dependence of the $ML$ relations on the chemical composition. The above relations are obtained for the entire set of BL~Her models with the mass range $0.5-0.8 M_{\odot}$. We also tested for the subset of BL~Her models with the mass range $0.5-0.6 M_{\odot}$ and found similar results indicating a dependence of the $ML$ relation on metallicity. However, we note here that the $MLZ$ relations exhibit quite a large scatter.

\section{Fourier parameters}
\label{sec:FP}

\begin{figure*}
\centering
\includegraphics[scale = 0.95]{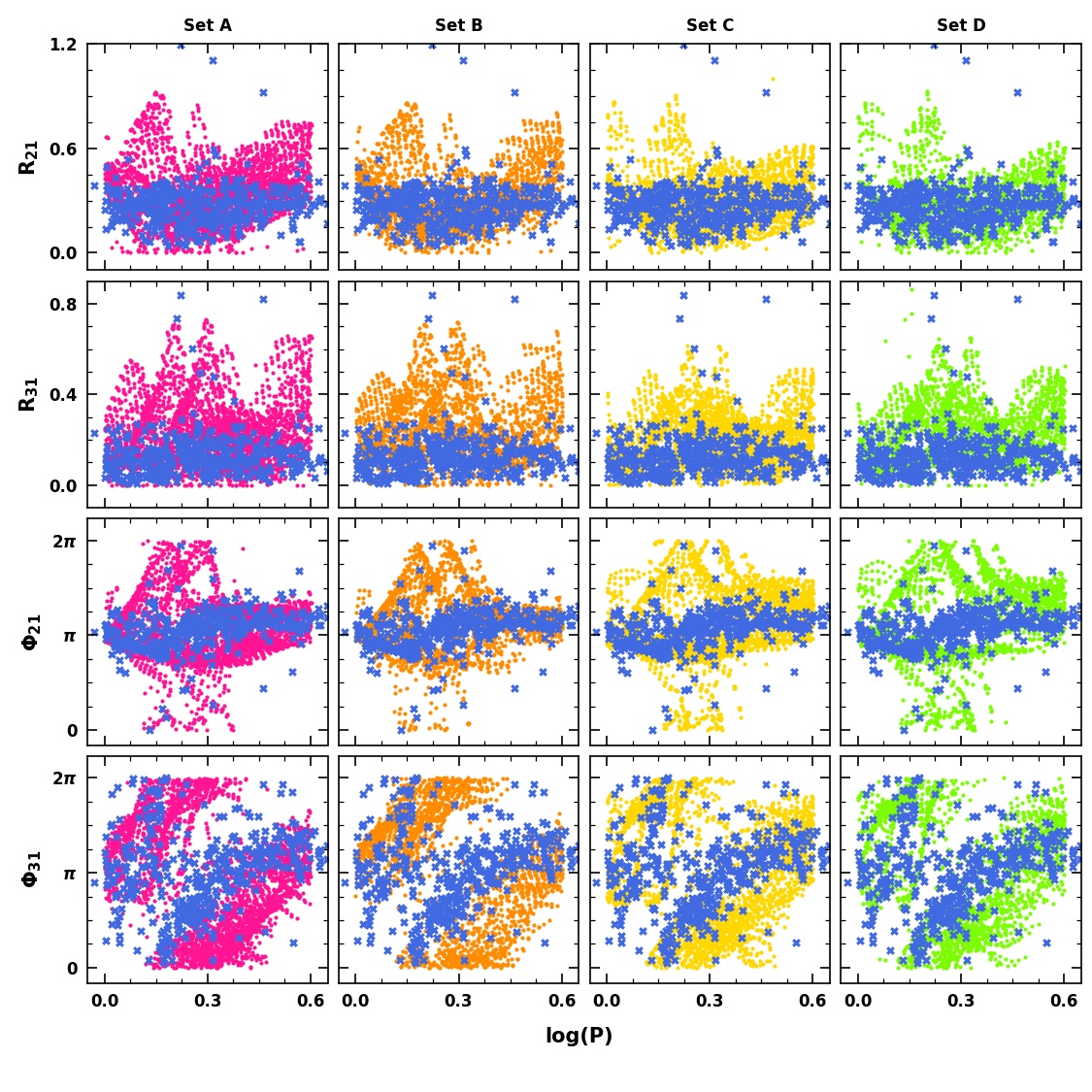}
\caption{Variation in the Fourier parameters of BL~Her models in the $G$ band as a function of period and the convective parameter sets~A (pink), B (orange), C (yellow), and D (green). Fourier parameters from the BL~Her stars in the All Sky region from Gaia DR3 are over-plotted (in navy crosses) for an easier comparison with the theoretical Fourier parameter space.}
\label{fig:FP_G}
\end{figure*}

\begin{figure*}
\centering
\includegraphics[scale = 0.95]{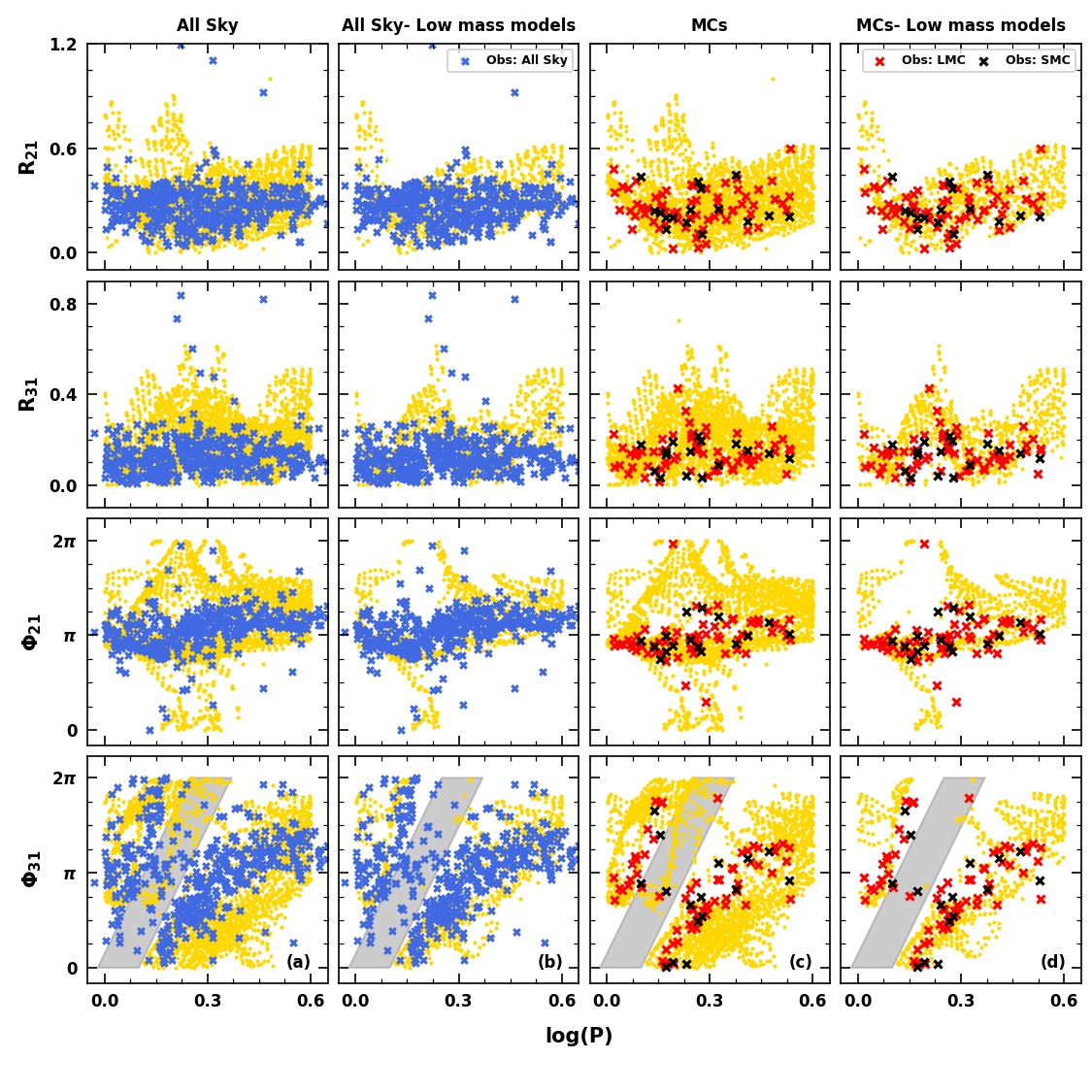}
\caption{Variation in the Fourier parameters of BL~Her models (displayed in yellow) in the $G$ band computed using the convection parameter set~C as a function of stellar mass and period. Low mass models indicate models with $M \leq 0.6M_{\odot}$. The grey shaded regions in the lowermost panels highlight that while low-mass models may suffice to model BL~Her stars in the Magellanic Clouds (sub-plots c, d), higher mass models may be needed to reliably model BL~Her stars in the All Sky region (sub-plots a, b).}
\label{fig:FP_mass}
\end{figure*}

\begin{figure*}
\centering
\includegraphics[scale = 0.95]{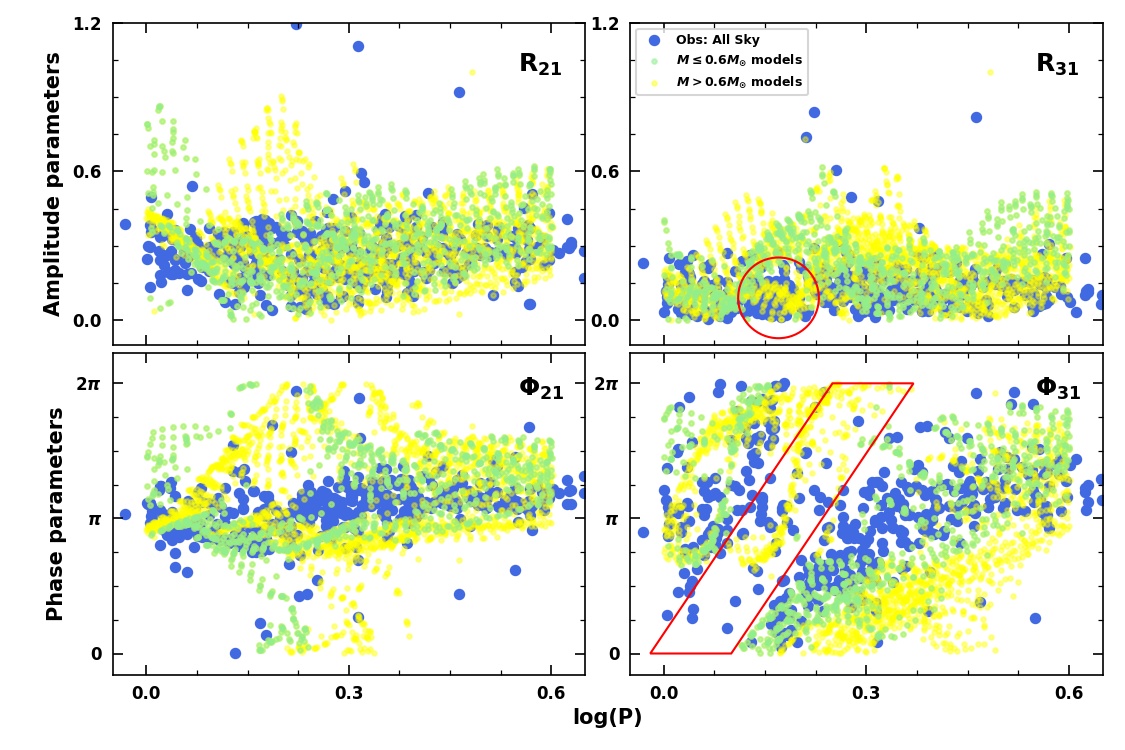}
\caption{Comparison of the theoretical Fourier parameters from the low-mass ($M \leq 0.6M_{\odot}$; green) and the high-mass models ($M > 0.6M_{\odot}$; yellow) with the observed Fourier parameters of BL~Her stars in the All~Sky region (displayed in navy). The regions marked with red highlight the presence of observed Fourier parameters but the lack of those from low-mass models.}
\label{fig:FP_mass_allsky}
\end{figure*}

The variation of the Fourier amplitude ($R_{21}$ and $R_{31}$) and phase ($\phi_{21}$ and $\phi_{31}$) parameters from the theoretical BL~Her light curves in the $G$ band is presented in Fig.~\ref{fig:FP_G} as a function of period and convection parameter sets. The observed $G$ band Fourier parameters from BL~Her stars in the All Sky region from $Gaia$ DR3 are also over-plotted. We find that the observed Fourier parameter space is well-covered by our BL~Her models. However, the models also produce Fourier amplitude parameters ($R_{21}$ and $R_{31}$) that are  considerably larger than what is observed. While this holds true for all the four sets of convection parameters, subtle differences exist among the four sets of convection parameters. For example, in the range of $0.4 < \log(P) < 0.5$, the mean values (with corresponding standard deviations on the mean value within the period bin) are 0.377$\pm$0.123 for set~A, 0.361$\pm$0.138 for set~B, 0.314$\pm$0.119 for set~C and 0.280$\pm$0.118 for set~D. The corresponding values of $R_{31}$ are 0.196$\pm$0.104 for set~A, 0.213$\pm$0.117 for set~B, 0.167$\pm$0.090 for set~C, and 0.202$\pm$0.090 for set~D. The choice of convection parameters therefore clearly play an important role in the theoretical light curve structures, especially on the Fourier amplitude parameters \citep[also see ][]{kovacs2023}. Adopting a higher mixing length (see e.g. \citealt{marconi2005,marconi2007,netzel2023} for RR Lyraes, \citealt{marconi2013b,marconi2013a} for classical Cepheids and review by \citealt{joyce2023a}) or different opacity values \citep{kanbur2018} may help resolve some of the discrepancies related to the Fourier amplitude parameters. We emphasize here that the convection parameter sets used in this analysis are the same as those outlined in \citet{paxton2019}, without any changes to the free convection parameters.

Since BL~Her stars are mostly considered to be low-mass stars, we also present the comparison of the Fourier parameters of the BL~Her stars from $Gaia$ DR3 with those from the subset of low-mass models ($0.5-0.6 M_{\odot}$) computed using the convection parameter set~C in Fig.~\ref{fig:FP_mass}. The theoretical Fourier parameters are displayed in yellow circles while the observed Fourier parameters from the All~Sky and the Magellanic Clouds are shown in navy, red, and black crosses, respectively.
We find two interesting results. For the Magellanic Clouds, the observed $G$-band Fourier parameter space for BL~Her stars is still covered well by the theoretical Fourier amplitude ($R_{21}$ and $R_{31}$) and phase ($\phi_{21}$ and $\phi_{31}$) parameter space of the low-mass models. In contrast, we observe a lack of models corresponding to a subset of BL~Her stars in the All Sky region (see, for example, the shaded region displaying the observed $\phi_{31}$ parameters of BL~Her stars in the All Sky region around $\log(P)=0.15$). This could indicate the need for higher mass models ($> 0.6 M_{\odot}$) to reliably model the observed light curves of BL~Her stars in the All Sky region.

We probe this further in Fig.~\ref{fig:FP_mass_allsky}, where we compare the Fourier parameters from observed BL~Her stars in the All~Sky region with those from low-mass ($M \leq 0.6M_{\odot}$) and high-mass models ($M > 0.6M_{\odot}$). The regions bounded by red in Fig.~\ref{fig:FP_mass_allsky} highlight the lack of low-mass models not only with respect to the $\phi_{31}$ parameter, but also to\ the $R_{31}$ parameter. These regions exhibit the presence of observed Fourier parameters from the All~Sky BL~Her stars, as well as theoretical Fourier parameters from BL~Her models with stellar mass $M > 0.6M_{\odot}$. However, further investigations using a more rigorous comparison of theoretical-observed light curve pairs are required to support our claim for the need of higher mass models corresponding to BL~Her stars in the All-Sky region.

\section{Summary and conclusion}
\label{sec:results}

We extended the fine grid of BL~Her models computed using \textsc{mesa-rsp}, the state-of-the-art 1D stellar pulsation code \citep{paxton2019} in \citetalias{das2021} to obtain theoretical light curves in the $Gaia$ passbands ($G,\,G_{\rm BP},\,G_{\rm RP}$) and carried out a comparative analysis with the observed light curves of BL~Her stars from $Gaia$ DR3. The grid of models encompasses a wide range of input parameters: metallicity ($-2.0\; \mathrm{dex} \leq \mathrm{[Fe/H]} \leq 0.0\; \mathrm{dex}$)\footnote{For equivalent $Z$ range, see Table~\ref{tab:composition}}, stellar mass (0.5--0.8\,$M\,_{\odot}$), stellar luminosity (50--300\,$L_{\odot}$), and effective temperature (full extent of the instability strip in steps of 50\,K; 5900--7200\,K for 50\,$L_{\odot}$ and 4700--6550\,K for 300\,$L_{\odot}$) and is computed using the four sets of convection parameters as outlined in \citet{paxton2019}. The input stellar parameters ($ZXMLT_{\rm eff}$) chosen are typical for BL~Her stars \citep[see e.g. ][]{buchler1992, marconi2007, smolec2016}. The BL~Her models we selected to use in this analysis exhibit positive growth rates in the fundamental mode with pulsation periods between one and four days and fulfill the condition of full-amplitude stable pulsations. 

In this way, we derived the theoretical $PL$ and $PW$ relations for these models in the $Gaia$ passbands and studied the effect of metallicity and convection parameters on these relations at mean light. We subsequently carried out a comparison of the light curve structures between the models and the  $Gaia$ DR3 data of BL~Her stars with respect to their Fourier parameters. The important results from this analysis are summarized below:

\begin{enumerate}
  \item We found statistically similar $PL$ and $PW$ slopes as a function of convection parameters across all $Gaia$ passbands for the complete set of BL~Her models computed using radiative cooling (sets B and D).

    \item The All-Sky BL~Her stars exhibit statistically different $PL$ slopes compared to the theoretical $PL$ slopes computed using the four sets of convection parameters. However, the empirical $PW$ relations from the All~Sky BL~Her stars mostly exhibit statistically similar slopes with the theoretical $PW$ slopes computed using the four sets of convection parameters, with a few exceptions.

    \item The empirical $PL$ and $PW$ slopes from the Magellanic  Cloud BL~Her stars are mostly statistically similar with the theoretical relations computed using the different convection parameter sets across all the $Gaia$ passbands.

   \item The theoretical $PL$ relations in the $Gaia$ passbands for BL~Her models exhibit steeper slopes and smaller dispersion with increasing wavelengths ($G_{BP} < G < G_{RP}$). This result is consistent with the theoretical relations for the Johnson-Cousin-Glass bands ($UBVRIJHKLL'M$) in \citetalias{das2021}.

    \item For both the complete set of models and the low-mass models only, we find the metallicity contribution to the $PLZ$ relations to be negligible within 3$\sigma$ uncertainties, except for a few cases using $G_{RP}$. 
    
    \item We find a small but significant negative coefficient of metallicity in the $PWZ$ relations for the BL~Her models using the four sets of convection parameters.

     \item The small but significant metallicity contribution in the Wesenheit relations arises from the high metallicity BL~Her models and could be a result of the increased sensitivity of bolometric corrections to metallicities at wavelengths shorter than the $V$ band \citep{gray2005, kudritzki2008}.

     \item The computed BL~Her models suggest a dependence of the $ML$ relation on chemical composition. This is in agreement with \citet{bono2020} where they found a marginal dependence of the $ML$ relation of T2Cs on metallicity.

    \item The observed Fourier parameter space is well-covered by our BL~Her models. However, the models have much larger Fourier amplitude parameters ($R_{21}$ and $R_{31}$) than what is observed. 

    \item For the Magellanic Clouds, the observed Fourier parameter space for BL~Her stars is still covered well by the Fourier parameter space of the low-mass models ($\leq 0.6 M_{\odot}$). Higher mass models ($> 0.6 M_{\odot}$) may be needed to reliably model the observed light curves of BL~Her stars in the All Sky region. However, further investigation is required to support this claim.
\end{enumerate}

In this work, we have used the \textsc{mesa-rsp} code, which computes non-linear radial stellar pulsations using the input stellar parameters ($ZXMLT_{\rm eff}$) typical for BL~Her stars. As noted in \citet{smolec2008}, these input parameters can be chosen without being bound by evolutionary tracks. The recent work by \citet{bono2020} offers us an opportunity to compute evolutionary tracks of T2Cs using the evolutionary code of \textsc{mesa} and compare the two different approaches. Although this is beyond the scope of the present study, it is indeed interesting for a future project.

Our grid of BL~Her models ushers in the era of large number statistics on the theoretical front in the study of stellar pulsations. However, note that this analysis uses the four convection parameter sets as outlined in \citet{paxton2019}; these convection parameter values are merely useful starting choices. Our grid of models suggests that the choice of convection parameters clearly play an important role in the theoretical light curve structures, especially on the Fourier amplitude parameters and therefore, may be reasonably changed to obtain better fitted models for a particular observed light curve. As an application of our extensive light curve analysis, we compared the theoretical and observed light curves of the BL~Her stars in the LMC with similar periods and obtained the best-fit models in an ongoing project (Das et al., Paper~III, in prep.). We note here that the theoretical Fourier parameter space from our present grid of BL~Her models is much larger than the observed Fourier parameter space. As an inverse method of stellar population synthesis, a comparison between the theoretical $PL$ and $PW$ relations obtained from the subset of the best-matched observed-model pairs for the BL~Her stars in the LMC and the theoretical relations from the complete set of BL~Her models would thus be of critical importance.

The results demonstrate that the empirical $PL$ and $PW$ relations from the BL~Her stars in the Magellanic Clouds and show statistically similar slopes with those from models across the four different sets of convection parameters; however, the theoretical slopes from the different convection sets themselves exhibit statistical differences amongst each other indicate that the observations are not yet precise enough to distinguish among the models with different convection parameter sets. It is also important to note that the theoretical and the empirical $PL$ and $PW$ relations have been compared at mean light, which exhibits a behaviour averaged over a pulsation cycle. In the era of percent-level precision, it is therefore important to compare the observations and the stellar pulsation models not just at mean light but above and beyond with respect to their light curve structures over a complete pulsation cycle. In light of this, we also plan to carry out a multiphase study \citep[for classical Cepheids, see][]{kurbah2023} for our grid of BL~Her models in the near future.

\begin{acknowledgements}
The authors thank the referee for useful comments and suggestions that improved the quality of the manuscript. S. D. expresses her heartfelt gratitude to Zs\'ofia Fejes (HR generalist, CSFK) for her continuous support throughout that made this project work smoother. S. D.\ and L. M.\ acknowledge the KKP-137523 `SeismoLab' \'Elvonal grant of the Hungarian Research, Development and Innovation Office (NKFIH). M. J. gratefully acknowledges funding from MATISSE: \textit{Measuring Ages Through Isochrones, Seismology, and Stellar Evolution}, awarded through the European Commission's Widening Fellowship. This project has received funding from the European Union's Horizon 2020 research and innovation program. RS is supported by the Polish National Science Centre, SONATA BIS grant, 2018/30/E/ST9/00598. The authors acknowledge the use of High Performance Computing facility Pegasus at IUCAA, Pune as well as at the HUN-REN (formerly ELKH) Cloud and the following software used in this project: \textsc{mesa}~r11701 \citep{paxton2011, paxton2013, paxton2015, paxton2018, paxton2019}. This research has made use of the Spanish Virtual Observatory (https://svo.cab.inta-csic.es) project funded by MCIN/AEI/10.13039/501100011033/ through grant PID2020-112949GB-I00. This research has made use of NASA’s Astrophysics Data System.

\end{acknowledgements}

\bibliographystyle{aa} 

\begin{appendix}
\section{Number of BL~Her models computed}

A summary of the number of BL~Her models in each convection parameter set used in the analysis is presented in Table~\ref{tab:number}. We also refer to Section~2 of \citetalias{das2021} for more details.

\begin{table*}[h!]

\caption{Summary of the number of BL~Her models in each convection parameter set used in the analysis.}
\centering \footnotesize
\scalebox{1}{
\begin{tabular}{l c c c c}
\hline\hline
Condition & Set A & Set B & Set C & Set D\\
\hline \hline
Total $ZXMLT_{\rm{eff}}$ combinations & 20412 & 20412 & 20412 & 20412\\
(Models computed in the linear grid) & & & & \\
\hline
Models with positive growth rate of the F-mode & 4481 & 4356 & 4061 & 4192\\
and with linear period: $0.8 \leq P \leq 4.2$ & & & & \\
(Models computed in the non-linear grid) & & & & \\
\hline
Models with non-linear period: $1 \leq P \leq 4$ & 4049 & 3854 & 3629 & 3678\\
\hline
Models with full-amplitude stable pulsation$^\dagger$ &3266  &2260  &2632  &2122\\
\hline
\end{tabular}
}
\tablefoot{
        \small $^{\dagger}$ Satisfies the condition that the amplitude of radius variation $\Delta R$, period $P$ and fractional growth rate $\Gamma$ do not vary by more than 0.01 over the last $\sim$100-cycles of the total 4000-cycle integrations. For a clear, pictorial representation of full-amplitude stable pulsation, the reader may refer to Fig. 2 of \citetalias{das2021}.}
\label{tab:number}
\end{table*}

\FloatBarrier

\section{Theoretical $PL$ and $PW$ slopes as a function of chemical composition}
\begin{figure*}
\centering
\includegraphics[scale = 0.95]{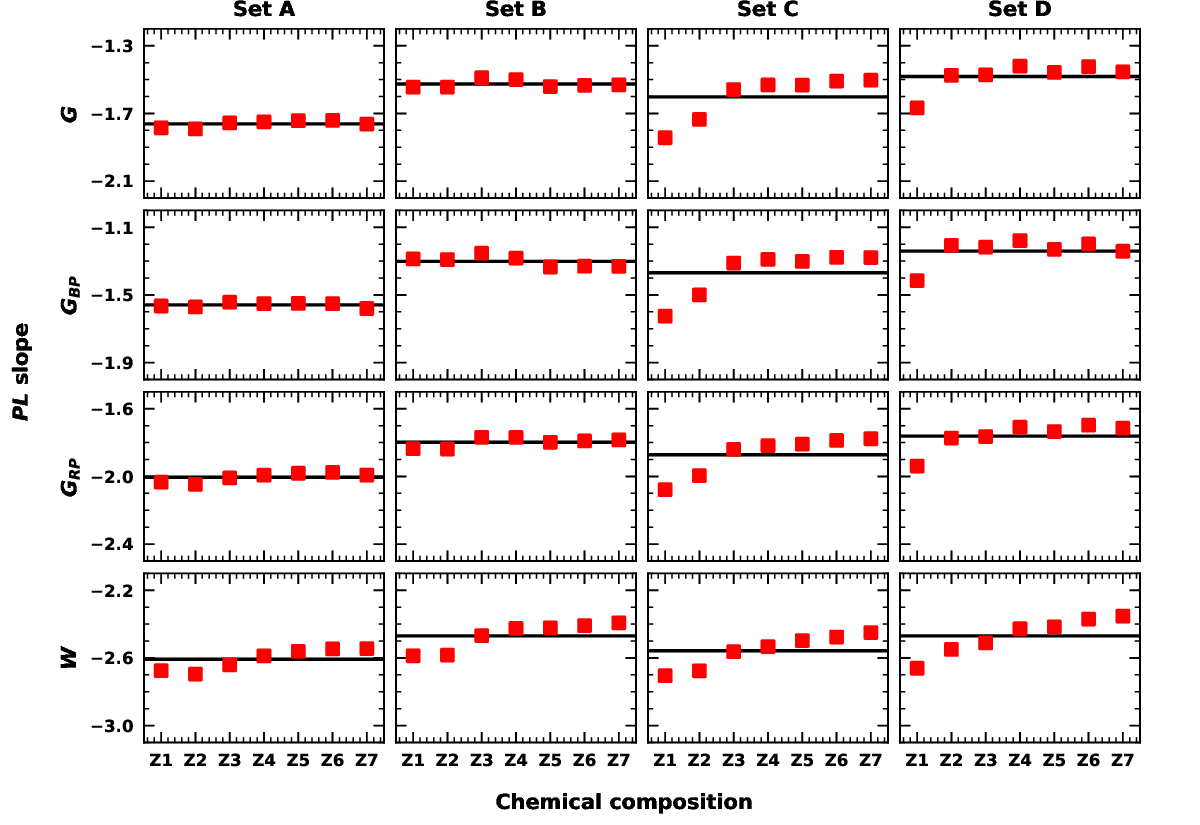}
\caption{Comparison of the $PL$ and $PW$ slopes of the BL~Her models as a function of chemical composition ($Z1=0.01300, Z2=0.00834, Z3=0.00424, Z4=0.00135, Z5=0.00061, Z6=0.00043, Z70.00014$) across different wavelengths for the convective parameter sets~A, B, C, and D. The horizontal lines represent the mean values of the slopes in the individual sub-plots. The y-scale is same (1 unit) in each panel for a relative comparison.}
\label{fig:PL_diffZ}
\end{figure*}

A comparison of the $PL$ and $PW$ slopes of the BL~Her models as a function of chemical composition across different wavelengths for the four sets of convective parameters is  presented in Fig.~\ref{fig:PL_diffZ}.

\FloatBarrier
\section{Comparison of the theoretical $PL$ and $PW$ relations using linear and non-linear pulsation periods}

A comparison of the theoretical $PL$ and $PW$ relations for BL~Her models of the mathematical form $M_\lambda=a\log(P)+b$ using linear and non-linear pulsation periods is presented in Table~\ref{tab:PL_linear}.

\begin{table*}
\caption{Comparison of the theoretical $PL$ and $PW$ relations for BL~Her models of the mathematical form $M_\lambda=a\log(P)+b$ for the complete set of models using linear and non-linear pulsation periods. $N$ is the total number of models.}
\centering
\scalebox{1}{
\begin{tabular}{c c c c c c c c c c}
\hline\hline
Band & Source & $a_{\rm NL}$ & $b_{\rm NL}$ & $\sigma_{\rm NL}$ & $N_{\rm NL}$ & $a_{\rm L}$ & $b_{\rm L}$ & $\sigma_{\rm L}$ & $N_{\rm L}$\\
\hline \hline
& & \multicolumn{4}{c}{Using non-linear periods} & \multicolumn{4}{c}{Using linear periods}\\
\hline
$G$ & $\rm{Z_{all}}$ (Set A) &-1.76$\pm$0.029&-0.263$\pm$0.01&0.262&3266&-1.747$\pm$0.03&-0.272$\pm$0.01&0.264&3266\\
$G$ & $\rm{Z_{all}}$ (Set B) &-1.531$\pm$0.031&-0.261$\pm$0.01&0.237&2260&-1.517$\pm$0.031&-0.268$\pm$0.01&0.239&2260\\
$G$ & $\rm{Z_{all}}$ (Set C) &-1.63$\pm$0.032&-0.171$\pm$0.011&0.267&2632&-1.616$\pm$0.032&-0.179$\pm$0.011&0.269&2632\\
$G$ & $\rm{Z_{all}}$ (Set D) &-1.49$\pm$0.033&-0.183$\pm$0.011&0.255&2122&-1.495$\pm$0.033&-0.182$\pm$0.011&0.253&2122\\
\hline
$G_{BP}$ & $\rm{Z_{all}}$ (Set A) &-1.543$\pm$0.033&-0.077$\pm$0.011&0.293&3266&-1.527$\pm$0.033&-0.086$\pm$0.011&0.295&3266\\
$G_{BP}$ & $\rm{Z_{all}}$ (Set B) &-1.296$\pm$0.034&-0.068$\pm$0.011&0.262&2260&-1.282$\pm$0.034&-0.074$\pm$0.011&0.263&2260\\
$G_{BP}$ & $\rm{Z_{all}}$ (Set C) &-1.382$\pm$0.036&0.034$\pm$0.012&0.299&2632&-1.367$\pm$0.036&0.027$\pm$0.012&0.3&2632\\
$G_{BP}$ & $\rm{Z_{all}}$ (Set D) &-1.238$\pm$0.037&0.024$\pm$0.013&0.284&2122&-1.245$\pm$0.036&0.026$\pm$0.012&0.282&2122\\ 
\hline
$G_{RP}$ & $\rm{Z_{all}}$ (Set A) &-2.015$\pm$0.025&-0.588$\pm$0.009&0.225&3266&-2.005$\pm$0.026&-0.596$\pm$0.009&0.228&3266\\
$G_{RP}$ & $\rm{Z_{all}}$ (Set B) &-1.81$\pm$0.027&-0.594$\pm$0.009&0.209&2260&-1.795$\pm$0.027&-0.601$\pm$0.009&0.21&2260\\
$G_{RP}$ & $\rm{Z_{all}}$ (Set C) &-1.909$\pm$0.028&-0.523$\pm$0.01&0.231&2632&-1.896$\pm$0.028&-0.531$\pm$0.01&0.233&2632\\
$G_{RP}$ & $\rm{Z_{all}}$ (Set D) &-1.779$\pm$0.029&-0.536$\pm$0.01&0.223&2122&-1.78$\pm$0.029&-0.536$\pm$0.01&0.221&2122\\
\hline
$W(G,G_{BP}-G_{RP})$ & $\rm{Z_{all}}$ (Set A) &-2.656$\pm$0.018&-1.234$\pm$0.006&0.159&3266& -2.654$\pm$0.018&-1.242$\pm$0.006&0.162&3266\\
$W(G,G_{BP}-G_{RP})$ & $\rm{Z_{all}}$ (Set B) &-2.507$\pm$0.021&-1.261$\pm$0.007&0.159&2260& -2.491$\pm$0.021&-1.27$\pm$0.007&0.161&2260\\
$W(G,G_{BP}-G_{RP})$ & $\rm{Z_{all}}$ (Set C) &-2.633$\pm$0.019&-1.231$\pm$0.007&0.163&2632& -2.623$\pm$0.02&-1.239$\pm$0.007&0.166&2632\\
$W(G,G_{BP}-G_{RP})$ & $\rm{Z_{all}}$ (Set D) &-2.517$\pm$0.021&-1.247$\pm$0.007&0.165&2122& -2.51$\pm$0.021&-1.25$\pm$0.007&0.163&2122\\
\hline
\end{tabular}}
\label{tab:PL_linear}
\end{table*}

\end{appendix}
\end{document}